\documentclass[aip,jcp,reprint,amsmath,amssymb,floatfix,citeautoscript,nofootinbib]{revtex4-2}
\usepackage{times}
\usepackage{mathptmx}
\DeclareMathAlphabet{\mathcal}{OMS}{cmsy}{m}{n} 
\usepackage[T1]{fontenc}
\usepackage{algorithm}
\usepackage{algpseudocode}
\usepackage{bm}
\usepackage{amsmath}
\usepackage{amssymb}
\usepackage{xcolor}
\usepackage{graphicx}
\usepackage{hyperref}
\hypersetup{
    colorlinks=true,
    linkcolor=blue,
    urlcolor=blue,
    citecolor=blue,
}
\usepackage{cleveref}
\usepackage{braket}
\usepackage[version=4]{mhchem}
\usepackage{multirow}
\usepackage{booktabs}

\newcommand{\mi}{\mathrm{i}}
\newcommand{\me}{\mathrm{e}}

\definecolor{myred}{rgb}{0.8,0,0}
\newcommand{\myred}[1]{{#1}}

\begin{document}

    \title{Fast Generation of Pipek--Mezey Wannier Functions via the Co-Iterative Augmented Hessian Method}

    \author{Gengzhi Yang}
    \affiliation{Joint Center for Quantum Information and Computer Science, University of Maryland, College Park, Maryland, 20742}
    \affiliation{Department of Mathematics, University of Maryland, College Park, Maryland, 20742}

    \author{Hong-Zhou Ye}
    \email{hzye@umd.edu}
    \affiliation{Department of Chemistry and Biochemistry, University of Maryland, College Park, Maryland, 20742}
    \affiliation{Institute for Physical Science and Technology, University of Maryland, College Park, Maryland, 20742}

    \date{\today}

    \begin{abstract}
        We report a $k$-point extension of the second-order co-iterative augmented Hessian (CIAH) algorithm, termed $k$-CIAH, for Pipek--Mezey (PM) localization of Wannier functions (WFs).
        By exploiting an efficient evaluation of the Hessian--vector product, $k$-CIAH achieves $O(N_k^2 n^3)$ scaling in both CPU time and memory, matching that of previously reported first-order $k$-space approaches while improving upon the $O(N_k^3 n^3)$ scaling of $\Gamma$-point CIAH, where $N_k$ denotes the number of $k$-points sampling the first Brillouin zone and $n$ characterizes the unit-cell size.
        Benchmark calculations on a diverse set of solids---including insulators, semiconductors, metals, and surfaces---demonstrate the fast and robust convergence of $k$-CIAH-based PMWF optimization, which yields an overall computational efficiency approximately $2$--$3$-fold higher than first-order $k$-space methods and orders of magnitude higher than $\Gamma$-point CIAH for localizing $1000$--$5000$ orbitals.
        The quality of the resulting PMWFs is further validated by accurate electronic band structures obtained via PMWF-based Wannier interpolation.
    \end{abstract}

    \maketitle

    \section{Introduction}
    \label{sec:intro}







    Wannier functions~\cite{Wannier37PR,Kohn59PR} (WFs) provide a localized real-space representation of Bloch orbitals and underpin a wide range of important applications, including band interpolation~\cite{Wang06PRB,Yates07PRB},
    evaluation of response properties~\cite{Giustino07PRB},
    Hamiltonian downfolding~\cite{Solovyev07PRB,Franchini12JPCM,Jiang23PRB,MosqueraLois24npjCM,Alvertis25PRApp},
    construction of machine-learning interatomic potentials~\cite{Zhang22JCP,Gao22NCommun,Gao24PCCP},
    and reduced-scaling many-body methods based on quantum embedding~\cite{Amadon08PRB,Cui20JCTC,Zhu20JCTC,Schafer21JCPa,Schafer21JCPb} and local correlation theories~\cite{Nejad25JCP,Zhu25JCP,Zhu25arXiv,Ye23arXiv,Ye24FD,Ye24JCTC,Hansen21MP}, among others.
    Several localization criteria have been developed, most prominently the Foster--Boys scheme~\cite{Foster60RMP} (commonly referred to as maximally localized WFs~\cite{Marzari97PRB,Souza01PRB,Marzari12RMP} in the physics and materials science communities), the fourth-moment scheme~\cite{Hoyvik12JCP}, the Edmiston--Ruedenberg scheme~\cite{Edmiston63RMP}, and the Pipek--Mezey (PM) scheme~\cite{Pipek89JCP}, which differ in how localization is quantified.

    Among these, PM localization is particularly attractive for periodic systems because it is formulated in terms of atomic populations, whose definition is straightforward under periodic boundary conditions, and yields chemically intuitive orbitals that preserve $\sigma$ and $\pi$ symmetry~\cite{Pipek89JCP}.
    The original PM formulation relies on Mulliken atomic populations~\cite{Pipek89JCP}, which are ill-defined in large basis sets with polarization and diffuse functions.
    This limitation has been largely overcome by the development of more robust population schemes, including meta-L\"{o}wdin~\cite{Sun14JCTC}, projection onto intrinsic atomic orbitals~\cite{Knizia13JCTC} or auxiliary minimal bases~\cite{Clement21JCTC}, and various real-space density partitioning approaches~\cite{Lehtola14JCTC}, thereby rendering PM-based localization reliable for calculations employing high-quality basis sets~\cite{Lehtola14JCTC,Ye23arXiv,Ye24FD,Ye24JCTC}.

    For periodic systems sampled with a uniform $k$-mesh, recent work has established gradient-based PMWF optimization in reciprocal space~\cite{Jonsson17JCTC,Clement21JCTC,Zhu24JPCA}, most notably through $k$-point implementations of the Broyden--Fletcher--Goldfarb--Shanno (BFGS) algorithm~\cite{Broyden1970,Fletcher1970,Goldfarb1970,Shanno1970}, which exhibit improved convergence compared to steepest-ascent or conjugate-gradient methods.
    These approaches generate PMWFs that preserve translational symmetry and avoid the cubic scaling of supercell-based $\Gamma$-point formulations, thereby enabling localization for increasingly large $k$-meshes and complex materials.
    Nevertheless, their performance remains limited by the intrinsic first-order convergence of quasi-Newton methods.

    In this work, we introduce a second-order strategy for PMWF generation based on the co-iterative augmented Hessian~\cite{Sun16arXiv} (CIAH) method.
    Originally developed for molecular orbital optimization~\cite{Sun16arXiv,Sun17CPL}, CIAH has been successfully applied to $\Gamma$-point supercell calculations in periodic solids~\cite{Ye23arXiv,Ye24FD,Ye24JCTC,Song25arXiv}.
    Here we generalize this framework to Bloch orbitals with $k$-point sampling, yielding what we term the $k$-CIAH method.
    By exploiting an efficient evaluation of the Hessian--vector product, $k$-CIAH achieves $O(N_k^2 n^3)$ scaling in both CPU time and memory, rivaling first-order $k$-space methods while retaining the quadratic convergence characteristic of molecular and $\Gamma$-point CIAH.
    Benchmark calculations on a diverse set of solids—including insulators, semiconductors, metals, and surfaces—demonstrate the fast and robust convergence of $k$-CIAH, resulting in overall computational efficiencies approximately $2$--$3$ times higher than the $k$-space BFGS algorithm and orders of magnitude higher than $\Gamma$-point CIAH when localizing $1000$--$5000$ orbitals.
    The quality of the resulting PMWFs is further validated by accurate electronic band structures obtained via PMWF-based Wannier interpolation~\cite{Yates07PRB,Marzari12RMP}.

    The rest of this paper is organized as follows.
    In \cref{sec:theory}, we present the theoretical framework of $k$-CIAH and its efficient implementation.
    \Cref{sec:comp_details} describes computational details.
    Numerical benchmarks, including convergence behavior (\cref{subsec:res_convergence}), cost analysis (\cref{subsec:res_cost}), and Wannier interpolation (\cref{subsec:res_wannier}), are reported in \cref{sec:results_and_discussion}, followed by concluding remarks in \cref{sec:conclusion}.

    \section{Theory}
    \label{sec:theory}

    \subsection{Notations}
    \label{subsec:notation}

    Throughout this paper, we consider a uniform $k$-point mesh $\mathcal{K}$ of size $N_k$ sampling the first Brillouin zone and $n_{\textrm{orb}}$ Bloch orbitals $\{\phi_{\bm{k}i}\}$ per $k$-point to be localized.
    \myred{The Bloch orbitals at each $k$-point are represented by their expansion coefficients in a set of $n_{\mathrm{AO}}$ translationally adapted Gaussian-type orbitals $\{\xi_{\bm{k}\rho}\}$ (henceforth referred to as AOs),
    \begin{equation}    \label{eq:phi_ki_AO_expansion}
        \phi_{\bm{k}i}(\bm{r})
            = \sum_{\rho}^{n_{\mathrm{AO}}} C_{\bm{k}\rho i} \xi_{\bm{k}\rho}(\bm{r}).
    \end{equation}
    The AOs at each $k$-point are non-orthogonal, giving rise to the overlap matrix,
    \begin{equation}
        S_{\bm{k}\rho\tau}
            = \braket{\xi_{\rho\bm{k}}|\xi_{\tau\bm{k}}}.
    \end{equation}}%
    Let $\bm{R}\in\mathcal{L}$ label the $N_k$ unit cells in the Born--von K\'{a}rm\'{a}n (BvK) supercell associated with $\mathcal{K}$.
    The Bloch orbitals can be transformed into an equal number of Wannier functions (WFs) in the BvK supercell via
    \begin{equation}
    \label{eq:w_Ri}
    w_{\bm{R}i}(\bm{r})
        = \sum_{\bm{k}}^{N_k} \phi_{\bm{k}i}(\bm{r})\, \theta_{\bm{R}\bm{k}}^* ,
    \end{equation}
    where $\theta_{\bm{R}\bm{k}}=(1/\sqrt{N_k})\,\me^{\mi\bm{R}\cdot\bm{k}}$.
    By construction, WFs in different unit cells are related by lattice translation,
    \begin{equation}
    \label{eq:w_Ri_w_0i}
    w_{\bm{R}i}(\bm{r})
        = w_{\bm{0}i}(\bm{r}-\bm{R}).
    \end{equation}

    Each unit cell contains $n_{\textrm{atom}}$ atoms labelled by $A$ and $n_{\textrm{proj}}$ atom-centered projectors $\{\chi_{\bm{k}\mu}\}$ per $k$-point, which are likewise Bloch functions \myred{and are represented by their AO expansion,
    \begin{equation}    \label{eq:chi_kmu_AO_expansion}
        \chi_{\bm{k}\mu}(\bm{r})
            = \sum_{\rho}^{n_{\mathrm{AO}}} D_{\bm{k}\rho \mu} \xi_{\bm{k}\rho}(\bm{r}).
    \end{equation}}%
    The corresponding Wannier-transformed projectors are defined analogously as
    \begin{equation}
    \label{eq:chi_Rmu}
    \chi_{\bm{R}\mu}(\bm{r})
        = \sum_{\bm{k}}^{N_k} \chi_{\bm{k}\mu}(\bm{r})\, \theta_{\bm{R}\bm{k}}^*
        \myred{= \sum_{\bm{k}}^{N_k} \sum_{\rho}^{n_{\mathrm{AO}}}
        D_{\bm{k}\rho, \bm{R}\mu} \xi_{\bm{k}\rho}(\bm{r})}
    \end{equation}
    and used to define the atomic populations entering the PM localization scheme described in \cref{subsec:PMWF}, \myred{where $D_{\bm{k}\rho,\bm{R}\mu} = D_{\bm{k}\rho\mu} \theta_{\bm{R}\bm{k}}^*$}.
    The size of the atomic projector basis depends on the projector type and typically ranges from that of a minimal basis to that of the full atomic-orbital basis.
    We note that $n_{\textrm{proj}}\ge n_{\textrm{orb}}$ is a necessary condition for obtaining well-defined atomic populations in \cref{eq:Q_TARi}.
    \myred{The notation for various symbols and their range is summarized in \cref{tab:notation}}.

    Within the BvK supercell, we denote the total numbers of atoms, AOs, atomic projectors, and orbitals to be localized by $N_{\textrm{atom}}$, $N_{\mathrm{AO}}$, $N_{\textrm{proj}}$, and $N_{\textrm{orb}}$, respectively; these are related to their per-cell or per-$k$-point counterparts by a factor of $N_k$ (e.g., $N_{\textrm{orb}} = N_k n_{\textrm{orb}}$).
    When analyzing computational scaling in \cref{subsec:cost}, we also use $n$ as a generic symbol for quantities that scale only with the unit-cell size (such as $n_{\textrm{orb}}$, $n_{\textrm{atom}}$, and $n_{\textrm{proj}}$).

    \begin{table}[!h]
        \centering
        \caption{Symbols and index ranges used throughout this work.}
        \label{tab:notation}
        \begin{tabular}{lcc}
            \toprule
            Quantity & Symbols & Range \\
            \midrule
            $k$-points & $\bm{k}, \bm{k}', \bm{q}$ & $N_k$ \\
            Unit cells & $\bm{R}, \bm{S}, \bm{T}$ & $N_k$ \\
            Atoms & $A$ & $n_{\mathrm{atom}}$ \\
            Bloch orbitals  & $i, j, m$ & $n_{\mathrm{orb}}$ \\
            Projectors & $\mu,\nu$ & $n_{\mathrm{proj}}$ \\
            Atomic orbitals & $\rho,\tau$ & $n_{\mathrm{AO}}$ \\
            \bottomrule
        \end{tabular}
    \end{table}

    \subsection{Pipek--Mezey Wannier functions (PMWFs)}
    \label{subsec:PMWF}

    The PMWFs are defined as WFs that maximize the PM objective function,
    \begin{equation}
    \label{eq:PM_obj}
    L_p
        = \sum_{\bm{T}A}^{N_{\textrm{atom}}}
          \sum_{i}^{n_{\textrm{orb}}}
          Q_{\bm{T}A,\bm{0}i}^{p} ,
    \end{equation}
    where $Q_{\bm{T}A,\bm{R}i}$ denotes the atomic population of the WF $w_{\bm{R}i}$ on atom $A$ in cell $\bm{T}$,
    \begin{equation}
    \label{eq:Q_TARi}
    Q_{\bm{T}A,\bm{R}i}
        = \braket{w_{\bm{R}i} | \hat{P}_{\bm{T}A} | w_{\bm{R}i}} ,
    \end{equation}
    and $p\ge 2$ is a positive integer.
    Exploiting the translational invariance of the WFs in \cref{eq:w_Ri_w_0i}, we include in \cref{eq:PM_obj} only the populations of WFs in a reference cell (taken as $\bm{R}=\bm{0}$).

    The atomic projection operator $\hat{P}_{\bm{T}A}$ is constructed from the Wannier-transformed projectors in \cref{eq:chi_Rmu}.
    For orthonormal projectors,
    \begin{equation}
    \label{eq:P_op_orth}
    \hat{P}_{\bm{T}A}
        = \sum_{\mu\in A}
          \ket{\chi_{\bm{T}\mu}}\bra{\chi_{\bm{T}\mu}} ,
    \end{equation}
    whereas for non-orthogonal projectors we use the symmetrized form
    \begin{equation}
    \label{eq:P_op_nonorth}
    \hat{P}_{\bm{T}A}
        = \frac{1}{2}\sum_{\mu\in A}
          \Big(
          \ket{\chi_{\bm{T}\mu}}\bra{\tilde{\chi}_{\bm{T}\mu}}
          +
          \ket{\tilde{\chi}_{\bm{T}\mu}}\bra{\chi_{\bm{T}\mu}}
          \Big) ,
    \end{equation}
    where
    \begin{equation}
    \label{eq:chitild_Tmu}
    \tilde{\chi}_{\bm{T}\mu}(\bm{r})
        = \sum_{\bm{R}\nu}^{N_{\textrm{atom}}}
          \chi_{\bm{R}\nu}(\bm{r})\,
          (O^{-1})_{\bm{R}\nu,\bm{T}\mu}
    \end{equation}
    are biorthogonal to $\chi_{\bm{T}\mu}$, with
    $O_{\bm{R}\mu,\bm{T}\nu}
     = \braket{\chi_{\bm{R}\mu}|\chi_{\bm{T}\nu}}$.
    The definitions in \cref{eq:P_op_orth,eq:P_op_nonorth} ensure that $\hat{P}_{\bm{T}A}$ is Hermitian, which in turn guarantees that $Q_{\bm{T}A,\bm{R}i}$ and the PM objective in \cref{eq:PM_obj} are real-valued even for complex WFs.

    The PM objective function is invariant under a translationally invariant (i.e.,~$\bm{R}$-independent) gauge transformation of the WFs,
    \begin{equation}
    \label{eq:w_Ri_gauge_symm}
    w_{\bm{R}i}(\bm{r})
        \rightarrow \me^{\mi\eta_i} w_{\bm{R}i}(\bm{r}),
    \qquad
    \forall\,\bm{R}\in\mathcal{L} .
    \end{equation}
    In reciprocal space, \cref{eq:w_Ri_gauge_symm} corresponds to a $k$-independent phase transformation of the Bloch orbitals,
    \begin{equation}
    \label{eq:phi_ki_gauge_symm}
    \phi_{\bm{k}i}(\bm{r})
        \rightarrow \me^{\mi\eta_i}\phi_{\bm{k}i}(\bm{r}),
    \qquad
    \forall\,\bm{k}\in\mathcal{K} .
    \end{equation}
    As discussed in \cref{subsec:parameterization_PMWFs}, fixing these gauge degrees of freedom reduces the number of independent parameters in the PMWF parameterization.

    \subsection{Parameterization of PMWFs}
    \label{subsec:parameterization_PMWFs}

    In this work, we parameterize the PMWFs in reciprocal space by applying unitary rotations to an initial set of Bloch orbitals $\{\phi_{\bm{k}i}^{(0)}\}$ (e.g., selected crystalline orbitals from a mean-field calculation),
    \begin{equation}
    \label{eq:w_Ri_from_Uk}
    w_{\bm{R}i}(\bm{r})
        = \sum_{\bm{k}}^{N_k}
          \left[
          \sum_{j}^{n_{\textrm{orb}}}
          \phi^{(0)}_{\bm{k}j}(\bm{r}) \, U_{\bm{k},ji}
          \right]
          \theta_{\bm{R}\bm{k}}^* .
    \end{equation}
    \Cref{eq:w_Ri_from_Uk} is equivalent to a real-space formulation in which a single, translationally invariant unitary transformation is applied to the initial WFs $\{w_{\bm{R}i}^{(0)}\}$ associated with $\{\phi_{\bm{k}i}^{(0)}\}$,
    \begin{equation}
    \label{eq:w_Ri_from_USR}
    w_{\bm{R}i}(\bm{r})
        = \sum_{\bm{S}j}^{N_{\textrm{orb}}}
          w^{(0)}_{\bm{S}j}(\bm{r}) \, U_{\bm{S}j,\bm{R}i} .
    \end{equation}
    The reciprocal- and real-space unitaries are connected by a double Fourier transform,
    \begin{equation}
    \label{eq:USR_from_Uk}
    U_{\bm{S}j,\bm{R}i}
        = \sum_{\bm{k}}
          \theta_{\bm{S}\bm{k}}\, U_{\bm{k},ji}\, \theta_{\bm{R}\bm{k}}^* ,
    \end{equation}
    from which it follows that the real-space unitary is translationally invariant,
    \begin{equation}
    \label{eq:USR_transl_inv}
    U_{\bm{S}j,\bm{R}i}
        = U_{\bm{0}j,(\bm{R}-\bm{S})i} .
    \end{equation}

    We parameterize each of the $N_k$ reciprocal-space unitary matrices in exponential form,
    \begin{equation}
    \label{eq:Uk_exp}
    U_{\bm{k}}
        = \me^{\kappa_{\bm{k}}} ,
    \end{equation}
    where the generators $\{\kappa_{\bm{k}}\}$ are anti-Hermitian,
    \begin{equation}
    \label{eq:Kk_antiHermi}
    \kappa_{\bm{k}}
        = -\kappa_{\bm{k}}^{\dagger} ,
    \end{equation}
    so that $U_{\bm{k}}$ is unitary by construction.
    Each $\kappa_{\bm{k}}$ contains $n_{\textrm{orb}}^{2}$ real degrees of freedom.
    Writing $\kappa_{\bm{k}} = X_{\bm{k}} + \mi Y_{\bm{k}}$ with real matrices $X_{\bm{k}}$ and $Y_{\bm{k}}$, a convenient choice of independent parameters is given by (i) the lower-triangular part of the antisymmetric matrix $X_{\bm{k}}$ (excluding the diagonal) and (ii) the lower-triangular part of the symmetric matrix $Y_{\bm{k}}$ (including the diagonal).

    To fix the gauge freedom identified in \cref{eq:phi_ki_gauge_symm}, we set $\mathrm{diag}(\kappa_{\bm{k}})=\mi\,\mathrm{diag}(Y_{\bm{k}})$ to zero for one chosen $k$-point.
    The total number of independent real parameters is therefore $N_k n_{\textrm{orb}}^{2}-n_{\textrm{orb}}$.
    These parameters generate $N_k$ independent complex unitaries $\{U_{\bm{k}}\}$, which correspond via \cref{eq:USR_from_Uk} to a single complex, translationally invariant real-space unitary acting on the supercell WFs.
    The special case in which this real-space unitary is constrained to be real-valued is discussed in \cref{subsec:real_rotations}.

    \subsection{Optimization of PMWFs using the co-iterative augmented Hessian (CIAH) method}
    \label{subsec:optimization_PMWFs}

    In this work, we determine the unitary rotations that transform the initial Bloch orbitals into the final PMWFs in \cref{eq:w_Ri_from_Uk} using the second-order co-iterative augmented Hessian (CIAH) algorithm~\cite{Sun16arXiv}.
    CIAH is a modified trust-region Newton method~\cite{Hoyvik12JCTC} that has been successfully applied to orbital localization in molecules~\cite{Sun16arXiv,Sun17CPL} and in periodic solids with $\Gamma$-point Brillouin-zone sampling~\cite{Ye23arXiv,Ye24FD,Ye24JCTC,Song25arXiv}.
    Near a local minimum, CIAH exhibits quadratic convergence while maintaining sufficient descent away from convergence~\cite{Sun16arXiv}.
    In this section, we extend the molecular and $\Gamma$-point formulations of \myred{the PM localization problem} to general Bloch orbitals and \myred{solve the resulting optimization problem using CIAH}; we refer to this application as $k$-CIAH.
    \myred{Thus, the term $k$-CIAH$,$ as used in this work, denotes the application of the generic CIAH optimizer to a PM objective function formulated for orbitals with $k$-point symmetry, rather than a modification of the underlying CIAH algorithm itself.}


    Starting from an initial guess $\{U_{\bm{k}}^{(0)}\}$ (see \cref{subsec:init_guess}), $k$-CIAH updates the $k$-space unitary rotations directly at each iteration,
    \begin{equation}
    \label{eq:Uk_update}
    U_{\bm{k}}^{(n+1)}
        = U_{\bm{k}}^{(n)} \me^{\kappa_{\bm{k}}^{(n+1)}} ,
    \end{equation}
    where the step in generator space $\{\kappa_{\bm{k}}^{(n+1)}\}$ is obtained by solving an augmented Hessian eigenvalue problem with the Davidson algorithm~\cite{Davidson75JCP},
    \begin{equation}
    \label{eq:AH_eigeqn}
    \begin{bmatrix}
    0 & \mathbf{g}^{(n)\dagger} \\
    \mathbf{g}^{(n)} & \mathbf{H}^{(n)}
    \end{bmatrix}
    \begin{bmatrix}
    1 \\ \mathbf{x}^{(n+1)}
    \end{bmatrix}
    = \epsilon
    \begin{bmatrix}
    1 \\ \mathbf{x}^{(n+1)}
    \end{bmatrix}.
    \end{equation}
    Here, $\mathbf{x}^{(n+1)}$ collects the independent real parameters in the generators $\{\kappa_{\bm{k}}^{(n+1)}\}$.
    $\mathbf{g}^{(n)}$ and $\mathbf{H}^{(n)}$ denote the gradient and Hessian of the negative PM objective in \cref{eq:PM_obj}, evaluated at the current orbitals and at zero generator,
    \begin{equation}    \label{eq:g_H_def}
    \begin{split}
    g^{(n)}_{\alpha}
        &= -\left.
           \frac{\partial L_p^{(n)}}{\partial x_{\alpha}}
           \right|_{\mathbf{x}=\mathbf{0}}, \\
    H^{(n)}_{\alpha\beta}
        &= -\left.
           \frac{\partial^2 L_p^{(n)}}{\partial x_{\alpha}\partial x_{\beta}}
           \right|_{\mathbf{x}=\mathbf{0}},
    \end{split}
    \end{equation}
    with $L_p^{(n)} = L_p[\{U_{\bm{k}}^{(n)}\}]$.
    \myred{We note that the optimization is formulated as a minimization problem in this work, which introduces the additional minus sign in \cref{eq:g_H_def}.}

    The analytical gradient, Hessian--vector product, and Hessian diagonal elements required to solve \cref{eq:AH_eigeqn} with the Davidson algorithm are derived in the Supporting Information.
    (The Hessian diagonals are used to precondition the Davidson update~\cite{Davidson75JCP}.)
    We summarize the working equations below.

    We first define two types of matrix elements of the atomic projection operators,
    \begin{equation}
    \label{eq:Pktij_P0kij}
    \begin{split}
    (P_{\bm{T}A,\bm{k}\bm{k}'})_{ij}
        &= \frac{1}{N_k}
           \braket{\phi_{\bm{k}i}|\hat{P}_{\bm{T}A}|\phi_{\bm{k}'j}}
         = \frac{1}{N_k}\sum_{\mu\in A}
           O_{\bm{T}\mu,\bm{k}i}^*\, O_{\bm{T}\mu,\bm{k}'j}, \\
    (P_{\bm{T}A,\bm{k}\bm{0}})_{ij}
        &= \frac{1}{\sqrt{N_k}}
           \braket{\phi_{\bm{k}j}|\hat{P}_{\bm{T}A}|w_{\bm{0}i}}
         = \frac{1}{\sqrt{N_k}}\sum_{\mu\in A}
           O_{\bm{T}\mu,\bm{k}i}^*\, O_{\bm{T}\mu,\bm{0}j},
    \end{split}
    \end{equation}
    where
    \begin{equation}
    \label{eq:O_Tmu_ki_0i}
    \begin{split}
    O_{\bm{T}\mu,\bm{k}i}
        &= \braket{\chi_{\bm{T}\mu}|\phi_{\bm{k}i}}
        = \sum_{\rho\tau}^{n_{\mathrm{AO}}^2}
        D_{\bm{k}\rho,\bm{T}\mu}^*
        S_{\bm{k}\rho\tau} C_{\bm{k}\tau i}, \\
    O_{\bm{T}\mu,\bm{0}i}
        &= \braket{\chi_{\bm{T}\mu}|w_{\bm{0}i}}
        = \sum_{\bm{k}} O_{\bm{T}\mu,\bm{k}i} \theta_{\bm{0}\bm{k}}^*
    \end{split}
    \end{equation}
    are overlaps between atomic projectors and Bloch orbitals (or WFs).
    At each $k$-point, the gradient corresponds to the lower-triangular part of the anti-Hermitian matrix
    \begin{equation}
    \label{eq:Gkij}
    G_{\bm{k}ij}
        = f_{ij}\,\hat{A}_{ij}
          \left\{
          -2p \sum_{\bm{T}A}^{N_{\textrm{atom}}}
          Q_{\bm{T}A,\bm{0}j}^{p-1}
          (P_{\bm{T}A,\bm{k}\bm{0}})_{ij}
          \right\},
    \end{equation}
    where $f_{ij}=1-\delta_{ij}/2$ and
    $\hat{A}_{ij}B_{\bm{k}ij}=(\mathbf{B}_{\bm{k}}-\mathbf{B}_{\bm{k}}^{\dagger})_{ij}$.
    Similarly, the Hessian--vector product at each $k$-point is given by the lower-triangular part of
    \begin{equation}
    \label{eq:Hvkij}
    (\mathbf{H}\mathbf{v})_{\bm{k}ij}
        = f_{ij}\,\hat{A}_{ij}
          \left\{
          \tilde{\sigma}^{\mathrm{d}}_{\bm{k}ij}
          + \tilde{\sigma}^{\mathrm{c\text{-}symm}}_{\bm{k}ij}
          + \tilde{\sigma}^{\mathrm{c\text{-}asymm}}_{\bm{k}ij}
          \right\},
    \end{equation}
    with
    \begin{equation}
    \label{eq:Hvp_d}
    \tilde{\sigma}^{\mathrm{d}}_{\bm{k}ij}
        = -4p(p-1)\sum_{\bm{T}A}^{N_{\textrm{atom}}}
          Q_{\bm{T}A,\bm{0}j}^{p-2}
          \sum_{\bm{k}'}^{N_k}
          \Re\!\left[(v_{\bm{k}'}^{\dagger}P_{\bm{T}A,\bm{k}'\bm{0}})_{jj}\right]
          (P_{\bm{T}A,\bm{k}\bm{0}})_{ij},
    \end{equation}
    \begin{equation}
    \label{eq:Hvp_csymm}
    \tilde{\sigma}^{\mathrm{c\text{-}symm}}_{\bm{k}ij}
        = -2p\sum_{\bm{T}A}^{N_{\textrm{atom}}}
          Q_{\bm{T}A,\bm{0}j}^{p-1}
          \sum_{\bm{k}'}^{N_k}
          (P_{\bm{T}A,\bm{k}\bm{k}'}v_{\bm{k}'})_{ij},
    \end{equation}
    \begin{equation}
    \label{eq:Hvp_casymm}
    \begin{split}
    \tilde{\sigma}^{\mathrm{c\text{-}asymm}}_{\bm{k}ij}
        &= -p\sum_{m}^{n_{\textrm{orb}}}
           \left[
           \sum_{\bm{T}A}^{N_{\textrm{atom}}}
           Q_{\bm{T}A,\bm{0}m}^{p-1}
           (P_{\bm{T}A,\bm{k}\bm{0}})_{im}
           \right] v_{\bm{k}jm}^* \\
        &\quad{}
           -p\sum_{\bm{T}A}^{N_{\textrm{atom}}}
           Q_{\bm{T}A,\bm{0}j}^{p-1}
           (v_{\bm{k}}^{\dagger}P_{\bm{T}A,\bm{k}\bm{0}})_{ij}.
    \end{split}
    \end{equation}
    These correspond to the disconnected, connected symmetric, and connected asymmetric contributions to the Hessian.
    Finally, the Hessian diagonal elements at each $k$-point are given by the lower-triangular part of the symmetric matrix
    \begin{equation}
    \label{eq:hkij}
    D_{\bm{k}ij}
        = f_{ij}\big(\tilde{D}_{\bm{k}ij}+\tilde{D}_{\bm{k}ji}\big),
    \end{equation}
    where the real and imaginary parts of $\tilde{\mathbf{D}}_{\bm{k}}$ read
    \begin{equation}
    \label{eq:hkij_ReIm}
    \begin{split}
    (\Re\tilde{\mathbf{D}}_{\bm{k}})_{ij}
        &= -4p(p-1)\sum_{\bm{T}A}^{N_{\textrm{atom}}}
           Q_{\bm{T}A,\bm{0}j}^{p-2}
           \big[(\Re P_{\bm{T}A,\bm{k}\bm{0}})_{ij}\big]^2 \\
        &\quad{}
           +2p\sum_{\bm{T}A}^{N_{\textrm{atom}}}
           Q_{\bm{T}A,\bm{0}i}^{p-1}
           \Re\!\left[
           (P_{\bm{T}A,\bm{k}\bm{0}})_{ii}
           -(P_{\bm{T}A,\bm{k}\bm{k}})_{jj}
           \right], \\
    (\Im\tilde{\mathbf{D}}_{\bm{k}})_{ij}
        &= -4p(p-1)\sum_{\bm{T}A}^{N_{\textrm{atom}}}
           Q_{\bm{T}A,\bm{0}j}^{p-2}
           \big[(\Im P_{\bm{T}A,\bm{k}\bm{0}})_{ij}\big]^2 \\
        &\quad{}
           +2p\sum_{\bm{T}A}^{N_{\textrm{atom}}}
           Q_{\bm{T}A,\bm{0}i}^{p-1}
           \Re\!\left[
           (P_{\bm{T}A,\bm{k}\bm{0}})_{ii}
           -(P_{\bm{T}A,\bm{k}\bm{k}})_{jj}
           \right].
    \end{split}
    \end{equation}

    We note that the analytical gradient in \cref{eq:Gkij} also enables PMWF optimization using gradient-based methods, such as the BFGS quasi-Newton algorithm~\cite{Broyden1970,Fletcher1970,Goldfarb1970,Shanno1970}, as explored in previous studies~\cite{Clement21JCTC,Zhu24JPCA}.
    In \cref{sec:results_and_discussion}, we compare the performance and computational efficiency of $k$-CIAH- and $k$-BFGS-based PMWF optimization.

    \subsection{Cost of $k$-CIAH-based PMWF optimization}
    \label{subsec:cost}

    The computational cost of $k$-CIAH-based PMWF optimization is dominated by repeated evaluations of the gradient in \cref{eq:Gkij} and the Hessian--vector product in \cref{eq:Hvkij} during the Davidson solution of the augmented Hessian eigenvalue problem in \cref{eq:AH_eigeqn}.
    \myred{The gradient evaluation involves three $O(N_k^2 n^3)$ steps:
    \begin{enumerate}
        \item $O(N_k^2 n_{\mathrm{AO}} n_{\mathrm{proj}} n_{\mathrm{orb}})$ CPU cost for computing the overlap matrix $O_{\bm{T}\mu,\bm{k}i}$ via \cref{eq:O_Tmu_ki_0i},
        \item $O(N_k^2 n_{\mathrm{proj}} n_{\mathrm{orb}}^2)$ CPU cost for constructing the projection matrix $(P_{\bm{T}A,\bm{k}\bm{0}})_{ij}$ via \cref{eq:Pktij_P0kij}, and
        \item $O(N_k^2 n_{\mathrm{atom}} n_{\mathrm{orb}}^2)$ CPU cost for evaluating the gradient via \cref{eq:Gkij}.
    \end{enumerate}
    Step 1 is $O(n_{\mathrm{AO}}/n_{\mathrm{orb}})$ more expensive than step 2, which is in turn $O(n_{\mathrm{proj}}/n_{\mathrm{atom}})$ more expensive than step 3.
    For typical applications, where $n_{\mathrm{AO}} \gg n_{\mathrm{orb}}$ and $n_{\mathrm{proj}} \gg n_{\mathrm{atom}}$, the CPU cost of the gradient evaluation is therefore dominated by step 1, i.e., the construction of the overlap matrix via \cref{eq:O_Tmu_ki_0i}.
    The memory cost is dominated by storage of the projection matrix $(P_{\bm{T}A,\bm{k}\bm{0}})_{ij}$, which scales as $O(N_k^2 n_{\mathrm{atom}} n_{\mathrm{orb}}^2) \sim O(N_k^2 n^3)$.
    A similar analysis shows that evaluation of the PM objective function [\cref{eq:PM_obj}] is likewise dominated by construction of the Wannier-transformed overlap matrix $O_{\bm{T}\mu,\bm{0}i}$ in \cref{eq:O_Tmu_ki_0i}, which also requires $O(N_k^2 n_{\mathrm{AO}} n_{\mathrm{proj}} n_{\mathrm{orb}}) \sim O(N_k^2 n^3)$ CPU cost.
    We therefore conclude that the function and gradient evaluations have the same leading CPU cost, arising from the evaluation of the overlap matrix via \cref{eq:O_Tmu_ki_0i}.
    }

    \myred{The Hessian--vector product requires the same intermediates $O_{\bm{T}\mu,\bm{k}i}$ and $(P_{\bm{T}A,\bm{k}\bm{0}})_{ij}$ and therefore has at least the same CPU and memory scalings as the gradient evaluation.
    Among the additional steps, the dominant cost arises from computing the projection--vector product,
    \begin{equation}
    \label{eq:Pv_naive}
        (Pv)_{\bm{T}A,\bm{k}ij}
            = \sum_{\bm{k}'}^{N_k}\sum_{m}^{n_{\textrm{orb}}}
              (P_{\bm{T}A,\bm{k}\bm{k}'})_{im}\, v_{\bm{k}'mj},
    \end{equation}
    which, as written, requires $O(N_k^3 n_{\mathrm{atom}} n_{\mathrm{orb}}^3) \sim O(N_k^3 n^4)$ CPU cost and $O(N_k^3 n_{\mathrm{atom}} n_{\mathrm{orb}}^2) \sim O(N_k^3 n^3)$ memory cost, both significantly higher than the $O(N_k^2 n^3)$ scaling of the gradient evaluation.
    However, by exploiting the factorized form of the projection operators in \cref{eq:Pktij_P0kij}, we can rewrite \cref{eq:Pv_naive} in the computationally more efficient form
    \begin{equation}
    \label{eq:Pv_fast}
        (Pv)_{\bm{T}A,\bm{k}ij}
            = \sum_{\mu\in A}
              O_{\bm{T}\mu,\bm{k}i}^*
              \left(
              \sum_{\bm{k}'}^{N_k}\sum_{m}^{n_{\textrm{orb}}}
              O_{\bm{T}\mu,\bm{k}'m}\, v_{\bm{k}'mj}
              \right).
    \end{equation}
    \Cref{eq:Pv_fast} completely avoids the unfavorable $O(N_k^3 n^3)$ storage associated with $(P_{\bm{T}A,\bm{k}\bm{k}'})_{ij}$ and reduces the CPU cost to $O(N_k^2 n_{\mathrm{proj}} n_{\mathrm{orb}}^2)$, which is comparable to that of step 2 above and therefore $O(n_{\mathrm{AO}}/n_{\mathrm{proj}})$ lower than the cost of building the overlap matrix in step 1.
    }

    \myred{With the projection--vector product evaluated efficiently via \cref{eq:Pv_fast}, the connected symmetric part [\cref{eq:Hvp_csymm}] can be computed with $O(N_k^2 n_{\mathrm{atom}} n_{\mathrm{orb}}^2)$ cost, comparable to step 3 above.
    The disconnected part [\cref{eq:Hvp_d}] can be evaluated at similar CPU cost by expressing the second term in terms of the projection--vector product,
    \begin{equation}
        \sum_{\bm{k}'}^{N_k}
        \Re\!\left[(v_{\bm{k}'}^{\dagger} P_{\bm{T}A,\bm{k}'\bm{0}})_{jj}\right]
        = \sum_{\bm{k}}^{N_k} (Pv)_{\bm{T}A,\bm{k}jj}.
    \end{equation}
    Finally, the connected asymmetric part [\cref{eq:Hvp_casymm}] can be evaluated with $O(N_k n_{\textrm{orb}}^3)$ cost as
    \begin{equation}
    \label{eq:Hvp_casymm_fast}
        \tilde{\sigma}^{\mathrm{c\text{-}asymm}}_{\bm{k}ij}
            = -p\sum_{m}^{n_{\textrm{orb}}}
              \left(
              A_{\bm{k}im}\, v_{\bm{k}jm}^*
              + v_{\bm{k}mi}^*\, A_{\bm{k}mj}
              \right),
    \end{equation}
    with intermediates
    \begin{equation}
    \label{eq:Akij}
        A_{\bm{k}ij}
            = \sum_{\bm{T}A}^{N_{\textrm{atom}}}
              Q_{\bm{T}A,\bm{0}j}\,(P_{\bm{T}A,\bm{k}\bm{0}})_{ij},
    \end{equation}
    which can be formed with $O(N_k^2 n_{\textrm{atom}} n_{\textrm{orb}}^2)$ CPU cost.
    }

    \myred{In summary, the CPU costs of the function, gradient, and Hessian--vector product evaluations are all dominated by the construction of the overlap matrices via \cref{eq:O_Tmu_ki_0i}, and therefore share the same leading $O(N_k^2 n_{\mathrm{AO}} n_{\mathrm{proj}} n_{\mathrm{orb}})$ scaling.
    In practice, however, the average cost of the Hessian--vector product is even lower and is primarily dominated by the $O(N_k^2 n_{\mathrm{proj}} n_{\mathrm{orb}}^2)$ cost of constructing the projection--vector product via \cref{eq:Pv_fast}.
    This is because the augmented Hessian [\cref{eq:AH_eigeqn}] remains fixed during its iterative diagonalization, so the expensive overlap matrices need to be computed only once per CIAH iteration.
    Table~S1 presents representative timing data that numerically validate this analysis.
    }

    In passing, we note that a real-space formulation of CIAH based on a supercell unitary rotation in \cref{eq:USR_from_Uk} can in principle achieve the same $O(N_k^2 n^3)$ scaling, provided that the translational invariance of the unitary in \cref{eq:USR_transl_inv} is explicitly enforced in the implementation.
    This is not the case, however, when one directly applies a molecular CIAH code to the BvK supercell with $\Gamma$-point Brillouin zone sampling, as in previous studies~\cite{Ye23arXiv,Ye24FD,Ye24JCTC,Song25arXiv}.
    In that setting, neglecting translational symmetry may break the translational structure of the initial Bloch orbitals (or WFs) and also increases the formal scaling to $O(N_k^3 n^3)$ in both CPU time and memory \myred{(see Supporting Information for details)}, making it computationally less favorable than the $k$-space approach developed here.
    We compare the two approaches numerically in \cref{sec:results_and_discussion}.
    \myred{A comparison of the computational cost of key steps in $k$-CIAH and $\Gamma$-CIAH is also provided in Table~S1.}

    \subsection{Comparison with $k$-BFGS-based PMWF optimization}
    \label{subsec:BFGS}

    \myred{The analytical gradient derived in \cref{subsec:optimization_PMWFs} also enables a $k$-BFGS algorithm that closely follows previous work~\cite{Clement21JCTC,Zhu24JPCA}, whose performance relative to $k$-CIAH will be examined in \cref{sec:results_and_discussion}.
    Our $k$-BFGS implementation shares the same overall framework as $k$-CIAH, but uses a modified update step [\textit{cf}.~\cref{eq:Uk_update}],
    \begin{equation}
    \label{eq:Uk_update_bfgs}
    U_{\bm{k}}^{(n+1)}
        = U_{\bm{k}}^{(n)} \me^{\alpha\, f_{\mathrm{cap}}(\kappa_{\bm{k}}^{(n+1)})} .
    \end{equation}
    Here, the search direction $\kappa_{\bm{k}}^{(n+1)}$ is determined using the ``two-loop recursion'' form of the limited-memory BFGS algorithm~\cite{Nocedal80MC} (which we refer to as BFGS henceforth), with the initial Hessian approximation taken to be the identity matrix.
    If the resulting direction is not a descent direction, we revert to the steepest-descent direction.
    To improve numerical stability, we further cap the search direction according to
    \begin{equation}
        f_{\mathrm{cap}}(\kappa)
            = \left\{
            \begin{split}
                \frac{s_0}{
                    \kappa_{\mathrm{max}}
                } \kappa,
                \qquad{}& \kappa_{\mathrm{max}} > s_0   \\
                \kappa,
                \qquad{}& \kappa_{\mathrm{max}} \leq s_0
            \end{split}
            \right.
    \end{equation}
    where $\kappa_{\mathrm{max}} = \max_{i} |\kappa_{i}|$ and $s_0 = 0.10$~a.u.~was found to provide good overall performance for the systems considered in this work (see Table~S2 in the Supporting Information for convergence benchmarks).
    The step length $\alpha$ is then determined by an Armijo backtracking line search following Zhu and Tew,~\cite{Zhu24JPCA} which requires only evaluations of the PM objective function [\cref{eq:PM_obj}].
    The history size, i.e.,~the number of previous iterations retained in the BFGS update, was set to $10$.
    We found the convergence to be relatively insensitive to this choice, consistent with the observations of Clement and co-workers.}~\cite{Clement21JCTC}

    \myred{Each $k$-BFGS iteration requires one gradient evaluation to update the approximate Hessian and at least two PM objective function evaluations for the Armijo line search, whose respective costs were discussed in \cref{subsec:cost}.
    The practical efficiency of $k$-BFGS relative to $k$-CIAH therefore depends on the total number of function and gradient evaluations required by $k$-BFGS, compared with the total number of gradient and Hessian--vector product evaluations required by $k$-CIAH.
    A detailed numerical comparison is presented in \cref{subsec:res_convergence}.}

    \subsection{Initial guess}
    \label{subsec:init_guess}

    We construct the initial guess $\{U_{\bm{k}}^{(0)}\}$ for PMWF optimization in two steps.
    First, an initial unitary $U_{\bm{k}_0}^{(0)}$ is obtained for a selected $k$-point $\bm{k}_0$ using either Cholesky decomposition~\cite{Aquilante06JCP} or an ``atomic'' projection guess~\cite{Sun16arXiv}, in which a set of projected atomic orbitals is constructed within the target subspace and used to initialize the localization~\cite{Pulay83CPL};
    the latter is the default initialization strategy in the \textsc{PySCF} code~\cite{Sun18WIRCMS,Sun20JCP} for molecular orbital localization.
    Second, for all $\bm{k}\neq \bm{k}_0$, we align the phase of the Bloch orbitals at $\bm{k}$ with those at $\bm{k}_0$ by setting
    \begin{equation}
    \label{eq:Uk0}
        U_{\bm{k}}^{(0)}
            = L_{\bm{k}}\, R_{\bm{k}}^{\dagger},
    \end{equation}
    where $L_{\bm{k}}$ and $R_{\bm{k}}$ are the left and right singular vectors from the singular value decomposition of
    \begin{equation}
    \label{eq:CkdaggerCk0}
        C_{\bm{k}}^{\dagger} C_{\bm{k}_0}\, U_{\bm{k}_0}^{(0)} ,
    \end{equation}
    with $C_{\bm{k}}$ the atomic-orbital coefficient matrix of $\{\phi_{\bm{k}i}^{(0)}\}$.
    As noted in prior work~\cite{Clement21JCTC,Zhu24JPCA}, the construction in \cref{eq:Uk0,eq:CkdaggerCk0} fixes the arbitrary gauge of the initial Bloch orbitals by aligning their phases to those at $\bm{k}_0$, which typically yields initial WFs that are primarily localized within each unit cell.

    \subsection{Real rotations}
    \label{subsec:real_rotations}

    In the special case where the $k$-point mesh $\mathcal{K}$ is closed under inversion,
    \begin{equation}
    -\bm{k}\in\mathcal{K}, \qquad \forall\,\bm{k}\in\mathcal{K},
    \end{equation}
    the real-space unitary rotation generated via \cref{eq:USR_from_Uk} can be chosen to be real-valued by imposing time-reversal symmetry (TRS) on the $k$-space generators,
    \begin{equation}
    \label{eq:Kk_TRS}
    \kappa_{\bm{k}} = \kappa_{-\bm{k}}^{*}.
    \end{equation}
    Throughout this section, $-\bm{k}$ is understood as $-\bm{k}\bmod\bm{G}$ for an appropriate reciprocal lattice vector $\bm{G}$ that maps $-\bm{k}$ back into the first Brillouin zone.

    The TRS constraint in \cref{eq:Kk_TRS} reduces the total number of independent parameters as follows.
    First, for the $N_k'\le N_k$ time-reversal invariant points in $\mathcal{K}$ (i.e., those satisfying $\bm{k}=-\bm{k}$), $\kappa_{\bm{k}}$ must be real and skew-symmetric and can therefore be parameterized by $n_{\textrm{orb}}(n_{\textrm{orb}}-1)/2$ real parameters.
    Second, the remaining $k$-points can be partitioned into $(N_k-N_k')/2$ time-reversal pairs $(\bm{k},-\bm{k})$ with $\bm{k}\ne -\bm{k}$; for each pair, only one generator needs to be parameterized explicitly, with its partner fixed by \cref{eq:Kk_TRS}.
    The total number of real parameters in the TRS case thus reduces to
    \begin{equation}
    \frac{1}{2}N_k' n_{\textrm{orb}}(n_{\textrm{orb}}-1)
    + \frac{1}{2}(N_k-N_k')n_{\textrm{orb}}^{2}
    = \frac{1}{2}\!\left(N_k n_{\textrm{orb}}^{2}-N_k' n_{\textrm{orb}}\right).
    \end{equation}

    The gradient and Hessian--vector product required for $k$-CIAH optimization under TRS can be obtained directly from their general forms in \cref{eq:Gkij,eq:Hvkij} by symmetrizing with respect to $\bm{k}\leftrightarrow -\bm{k}$,
    \begin{equation}
    \label{eq:Gkij_TRS}
    \overline{G}_{\bm{k}ij}
        = G_{\bm{k}ij}
        + (1-\delta_{\bm{k},-\bm{k}})\, G_{-\bm{k}ij}^{*},
    \end{equation}
    \begin{equation}
    \label{eq:Hvkij_TRS}
    (\overline{\mathbf{H}\mathbf{v}})_{\bm{k}ij}
        = (\mathbf{H}\mathbf{v})_{\bm{k}ij}
        + (1-\delta_{\bm{k},-\bm{k}})\,
          (\mathbf{H}\mathbf{v})_{-\bm{k}ij}^{*}.
    \end{equation}
    The Hessian diagonal elements are modified analogously from \cref{eq:hkij,eq:hkij_ReIm}.
    Specifically, $P_{\bm{T}A,\bm{k}\bm{0}}$ in \cref{eq:hkij_ReIm} is replaced by its TRS-packed form,
    \begin{equation}
    \overline{P}_{\bm{T}A,\bm{k}\bm{0}}
        = P_{\bm{T}A,\bm{k}\bm{0}}
        + (1-\delta_{\bm{k},-\bm{k}})\, P^*_{\bm{T}A,-\bm{k}\bm{0}},
    \end{equation}
    and $P_{\bm{T}A,\bm{k}\bm{k}}$ is replaced by
    \begin{equation}
    \label{eq:Pkk_TRS}
    \overline{P}_{\bm{T}A,\bm{k}\bm{k}}
        = P_{\bm{T}A,\bm{k}\bm{k}}
        + (1-\delta_{\bm{k},-\bm{k}})
          \big[
          P_{\bm{T}A,-\bm{k}(-\bm{k})}
          + P^*_{\bm{T}A,-\bm{k}\bm{k}}
          + P^*_{\bm{T}A,\bm{k}(-\bm{k})}
          \big].
    \end{equation}
    In practice, for all systems tested we found that replacing \cref{eq:Pkk_TRS} with the following diagonal approximation,
    \begin{equation}
    \label{eq:Pkk_TRS_diag}
    \overline{P}_{\bm{T}A,\bm{k}\bm{k}}
        \approx P_{\bm{T}A,\bm{k}\bm{k}}
        + (1-\delta_{\bm{k},-\bm{k}})\,
          P_{\bm{T}A,-\bm{k}(-\bm{k})},
    \end{equation}
    does not measurably affect the convergence rate of $k$-CIAH-based PMWF optimization, \myred{as reflected by the similar convergence behavior of $k$-CIAH and $\Gamma$-CIAH in \cref{subsec:res_convergence}, where the latter employs the full Hessian diagonals.}
    This is expected because the Hessian diagonals serve only as a preconditioner for solving the augmented Hessian eigenvalue problem in \cref{eq:AH_eigeqn}.
    All numerical results reported in \cref{sec:results_and_discussion} therefore employ the diagonal approximation in \cref{eq:Pkk_TRS_diag}.

    \subsection{Escaping from local stationary points}
    \label{subsec:stability_analysis}

    PMWF optimization can occasionally converge to a local minimum or a saddle point.
    The efficient Hessian--vector product developed in \cref{subsec:cost} enables a straightforward stability analysis: one can identify directions of negative curvature and escape saddle points by following the corresponding Hessian eigenvectors.
    In practice, however, our numerical experiments indicate that most instabilities are dominated by pairwise rotations between WFs, which can be treated effectively with a simple Jacobi sweep algorithm~\cite{Edmiston63RMP,Pipek89JCP,Schreder24JCP} which we describe in this section.

    Consider a real-space $2\times 2$ rotation that mixes a WF pair $(w_{\bm{0}i},w_{\bm{R}j})$ and all of its lattice translates by $\bm{R}_0 \in \mathcal{L}$,
    \begin{equation}
    \label{eq:Jacobi_real_space}
    \begin{bmatrix}
        \tilde{w}_{\bm{R}_0,i} &
        \tilde{w}_{\bm{R}_0+\bm{R},j}
    \end{bmatrix}
        =
        \begin{bmatrix}
            w_{\bm{R}_0,i} &
            w_{\bm{R}_0+\bm{R},j}
        \end{bmatrix}
        \begin{bmatrix}
            \cos\theta & \sin\theta \\
            -\sin\theta & \cos\theta
        \end{bmatrix}.
    \end{equation}
    In reciprocal space, \cref{eq:Jacobi_real_space} is equivalent to applying the following $k$-dependent rotation to the corresponding Bloch orbital pair,
    \begin{equation}
    \label{eq:Jacobi_k_space}
    \begin{bmatrix}
        \tilde{\phi}_{\bm{k}i} &
        \tilde{\phi}_{\bm{k}j}
    \end{bmatrix}
        =
        \begin{bmatrix}
            \phi_{\bm{k}i} &
            \phi_{\bm{k}j}
        \end{bmatrix}
        \begin{bmatrix}
            \cos\theta &
            \me^{\mi\bm{k}\cdot\bm{R}}\sin\theta \\
            -\me^{-\mi\bm{k}\cdot\bm{R}}\sin\theta &
            \cos\theta
        \end{bmatrix}.
    \end{equation}
    A Jacobi-sweep stability check can be constructed by examining all $N_k n_{\textrm{orb}}^2$ WF pairs $(w_{\bm{0}i},w_{\bm{R}j})$.
    This procedure scales as $O(N_k^2 n^3)$, comparable to a single gradient evaluation.
    In practice, the cost can be further reduced to $O(N_k n^3)$ by restricting $\bm{R}$ in \cref{eq:Jacobi_real_space,eq:Jacobi_k_space} to lie within a finite cutoff radius $R_{\textrm{max}}$, \myred{which we set to $10$~Bohr by default}.
    This truncation is well motivated because pairwise instabilities typically arise between WFs that are spatially close.
    The working equations for determining the optimal rotation angle $\theta$ and details of an efficient implementation are provided in the Supporting Information.

    \subsection{Wannier interpolation}
    \label{subsec:wannier_interpolation}

    An important application of \myred{maximally localized or Foster--Boys} WFs is the efficient calculation of electronic band structures via Wannier interpolation~\cite{Yates07PRB,Marzari12RMP}.
    The electronic energy bands $\{\epsilon_{\bm{k}i}\}$ are the eigenvalues of the Fock matrix $\{F_{\bm{k}ij}\}$ evaluated along a selected $k$-point path $\mathcal{K}_{\textrm{Band}}$, typically connecting high-symmetry points in the first Brillouin zone.
    In a conventional workflow, one first performs a self-consistent-field (SCF) calculation to obtain orbitals on a uniform $k$-point mesh $\mathcal{K}_{\textrm{SCF}}$, and then carries out a sequence of non-SCF calculations to construct and diagonalize the Fock matrix at $k$-points in $\mathcal{K}_{\textrm{Band}}$.
    For high-resolution band structures, this approach becomes computationally demanding because it requires a large number of Fock builds.

    Wannier interpolation circumvents this bottleneck by enabling efficient band-structure calculations within a chosen energy window.
    First, the SCF orbitals corresponding to the bands of interest are localized to yield WFs $\{w_{\bm{R}i}\}$, with $\{\phi_{\bm{k}i}\}$ denoting the associated Bloch orbitals.
    The real-space Fock matrix in the localized WF basis is then obtained by Fourier transformation,
    \begin{equation}
    \label{eq:FRij}
    F_{\bm{R}i,\bm{0}j}
        = \braket{w_{\bm{R}i}|\hat{F}|w_{\bm{0}j}}
        = \sum_{\bm{k}\in\mathcal{K}_{\textrm{SCF}}}
          \theta_{\bm{R}\bm{k}}\, F_{\bm{k}ij},
    \end{equation}
    where $F_{\bm{k}ij}=\braket{\phi_{\bm{k}i}|\hat{F}|\phi_{\bm{k}j}}$ can be obtained via basis transformation from the Fock matrix in the SCF orbital basis.
    Because localized WFs decay rapidly in real space~\cite{Kohn59PR,He01PRL}, one can approximate the Fock matrix at an arbitrary $k$-point $\bm{q}$ by truncating the real-space sum,
    \begin{equation}
    \label{eq:Fqij}
    F_{\bm{q}ij}
        \approx \tilde{F}_{\bm{q}ij}
        = \sum_{\bm{R}\in\mathrm{WS}(\mathcal{K}_{\textrm{SCF}})}
          \frac{1}{d_{\bm{R}}}\,
          \theta_{\bm{R}\bm{q}}^{*}\, F_{\bm{R}ij}.
    \end{equation}
    Here, $\bm{R}$ runs over the Wigner--Seitz cell associated with the BvK supercell of $\mathcal{K}_{\textrm{SCF}}$, and the factor $1/d_{\bm{R}}$ accounts for degeneracies at the Wigner--Seitz boundary~\cite{Yates07PRB,Mostofi08CPC,Marzari12RMP}.
    By construction, \cref{eq:Fqij} reproduces the exact SCF Fock matrix for $\bm{q}\in\mathcal{K}_{\textrm{SCF}}$.
    Diagonalizing $\tilde{F}_{\bm{q}ij}$ for $\bm{q} \in \mathcal{K}_{\textrm{Band}}$ yields the Wannier-interpolated band energies $\{\bar{\epsilon}_{\bm{q}i}\}$.
    We present numerical examples of Wannier interpolation based on PMWFs in \cref{subsec:res_wannier}.

    \section{Computational details}
    \label{sec:comp_details}

    \begin{table}[!h]
    \centering
        \caption{Information on the systems used in the numerical tests of this work.
        For the first six insulating or semiconducting systems, $n_{\textrm{orb}}$ denotes the number of occupied orbitals.
        For the remaining four metallic systems, Fermi smearing~\cite{Gillan89JPCM,Santos23PRB} with $\sigma=0.05$~eV is employed to facilitate SCF convergence; in these cases, $n_{\textrm{orb}}$ is chosen to be the maximum number of bands across all $k$-points with occupation greater than $10^{-6}$.}
        \label{tab:info}
        \begin{tabular}{llll}
            \toprule
            System & $k$-mesh & $n_{\textrm{atm}}$ & $n_{\textrm{orb}}$  \\
            \hline
            h-BN & $15 \times 15 \times 1$ & $2$ & $4$  \\
            Diamond & $7 \times 7 \times 7$ & $2$ & $4$  \\
            MgO & $7 \times 7 \times 7$ & $2$ & $8$  \\
            Silicon & $7 \times 7 \times 7$ & $2$ & $4$  \\
            \ce{SiO2} & $3 \times 3 \times 3$ & $9$ & $24$  \\
            CO/MgO(001) & $3 \times 3 \times 1$ & $18$ & $69$  \\
            \textit{trans}-$(\ce{C2H2})_{\infty}$ & $101 \times 1 \times 1$ & $4$ & $6$  \\
            C-nanotube & $11 \times 1 \times 1$ & $32$ & $65$  \\
            Graphene & $15 \times 15 \times 1$ & $2$ & $5$  \\
            Aluminum & $5 \times 5 \times 5$ & $4$ & $8$  \\
            \bottomrule
        \end{tabular}
    \end{table}

    We implemented the $k$-CIAH for PMWF optimization in a developer version of \textsc{PySCF}~\cite{Sun18WIRCMS,Sun20JCP}, which relies on \textsc{Libcint}~\cite{Sun15JCC} for evaluating Gaussian integrals.
    \myred{The core CIAH implementation for iterative diagonalization of the augmented Hessian originally implemented in PySCF for CIAH-based molecular PM localization~\cite{Sun16arXiv} is kept unchanged, and only the gradient and Hessian--vector product routines are redefined using the expressions derived in \cref{subsec:optimization_PMWFs}.
    All parameters originally introduced in Ref.~\onlinecite{Sun16arXiv} for CIAH-based molecular PM localization, including those used in the Davidson solver, are therefore adopted here directly without further reoptimization.}
    The orbitals to be localized are Kohn--Sham orbitals generated with the Perdew--Burke--Ernzerhof (PBE) exchange--correlation functional~\cite{Perdew96PRL} on uniform $k$-point meshes that include the $\Gamma$-point.
    Time-reversal symmetry is enforced in both the mean-field calculations and the PMWF optimization as described in \cref{subsec:real_rotations}, so that the resulting WFs are real-valued.
    All calculations employ Goedecker--Teter--Hutter (GTH) pseudopotentials optimized for PBE~\cite{Goedecker96PRB,Hartwigsen98PRB} together with the GTH-cc-pVDZ Gaussian basis sets~\cite{Ye22JCTC}.
    Range-separated density fitting~\cite{Ye21JCPa,Ye21JCPb} is used to construct the Coulomb matrix.

    PMWF optimizations are initialized using the atomic initial guess implemented in \textsc{PySCF} for the Bloch orbitals at the $\Gamma$-point, followed by the phase-alignment procedure described in \cref{subsec:init_guess}.
    The exponent $p$ is set to $2$ in all calculations.
    The optimization is deemed converged when the norm of the gradient falls below $10^{-5}$~a.u.~and that the PM objective value changes by less than $10^{-6}$~a.u.~between successive cycles.
    The atomic projectors are generated with the meta-L\"{o}wdin scheme~\cite{Sun14JCTC}.
    Our preliminary tests using alternative choices, such as intrinsic atomic orbitals~\cite{Knizia13JCTC}, yield comparable convergence behavior and qualitatively similar WFs to those obtained with meta-L\"{o}wdin projectors \myred{(Table~S4)}.
    This weak sensitivity to the choice of atomic projectors is consistent with previous reports~\cite{Lehtola14JCTC}.
    In all PMWF calculations, we perform Hessian- and Jacobi-sweep-based stability analyses after convergence.
    If an instability is detected, the resulting orbitals are used to reinitialize a subsequent PMWF optimization, which is repeated until a stable solution is obtained.
    All PMWF calculations were performed using 8 AMD EPYC 7763 CPU cores and a total of 32 GB of memory.

    In \cref{sec:results_and_discussion}, we compare the performance of $k$-CIAH with gradient-based $k$-BFGS (\cref{subsec:BFGS}) and with molecular CIAH applied directly to the BvK supercell without exploiting translational symmetry between unit cells (hereafter referred to as $\Gamma$-CIAH).
    \myred{The same phase-aligned atomic initial guess and convergence criteria as described above are used for $k$-BFGS and $\Gamma$-CIAH.}
    \Cref{tab:info} summarizes the 10 solid-state systems selected for numerical tests, spanning insulators, semiconductors, (semi)metals, and surfaces.
    The crystal structures are provided in the Supporting Information.
    \myred{For gapped systems, all occupied orbitals available in the employed pseudopotentials are localized, and $n_{\mathrm{orb}}$ is equal to the number of occupied orbitals per $k$-point.
    One exception is the band interpolation of h-BN discussed in \cref{subsec:res_wannier}, for which two virtual orbitals are localized together with the four occupied orbitals.
    For metallic systems, Fermi smearing with $\sigma = 0.05$~eV is used to facilitate SCF convergence, resulting in fractional occupations of the Bloch orbitals.
    In these cases, $n_{\textrm{orb}}$ is chosen as the maximum number of bands across all $k$-points with occupation greater than $10^{-6}$.
    We note that practical application of Wannier localization to metallic systems may benefit from a proper band-disentanglement procedure~\cite{Souza01PRB,Marzari12RMP,Damle18MMS}, which could serve as a preprocessing step before applying the PMWF optimization technique developed in this work.
    This is left for future work.
    }

    \begin{table*}[!t]
        \centering
        \caption{Convergence of PMWF optimization using $k$-CIAH, $k$-BFGS, and $\Gamma$-CIAH for selected systems.
        For each method, ``PM obj.''~denotes the converged PM objective value, $N_{\textrm{iter}}$ denotes the number of unitary updates defined in \cref{eq:Uk_update}, and $N_{\mathrm{f}+\mathbf{g}+\mathbf{Hv}}$ denotes the total number of PM objective function, gradient, and---for CIAH-based methods only---Hessian--vector product evaluations.
        For $k$-BFGS, instabilities are identified in the initially converged orbitals for MgO and are subsequently resolved using the stability analysis developed in \cref{subsec:stability_analysis}.
        In this case, $N_{\textrm{iter}}$ lists the iteration counts of the successive PMWF optimizations separated by ``$+$'', whereas $N_{\mathrm{f}+\mathbf{g}}$ reports the total number of evaluations accumulated over all optimizations.
        }
        \label{tab:conv}
        \begin{tabular}{llllllllll}
            \toprule
            \multirow{2}{*}{System}
            & \multicolumn{3}{c}{$k$-CIAH}
            & \multicolumn{3}{c}{$k$-BFGS}
            & \multicolumn{3}{c}{$\Gamma$-CIAH} \\
            \cmidrule(lr){2-4}
            \cmidrule(lr){5-7}
            \cmidrule(lr){8-10}
            & PM obj. & $N_{\textrm{iter}}$ & $N_{\mathrm{f}+\mathbf{g}+\mathbf{Hv}}$
            & PM obj. & $N_{\textrm{iter}}$ & $N_{\mathrm{f}+\mathbf{g}}$
            & PM obj. & $N_{\textrm{iter}}$ & $N_{\mathrm{f}+\mathbf{g}+\mathbf{Hv}}$  \\
            \midrule
            h-BN & 2.276 & $3$ & $56$ & 2.276 & $27$ & $80$ & 2.276 & $6$ & $70$  \\
            Diamond & 1.859 & $4$ & $81$ & 1.859 & $51$ & $152$ & 1.859 & $5$ & $77$  \\
            MgO & 3.241 & $9$ & $138$ & 3.241 & $19+5$ & $70$ & 3.241 & $9$ & $115$  \\
            Silicon & 1.798 & $5$ & $89$ & 1.798 & $49$ & $146$ & 1.798 & $5$ & $81$  \\
            \ce{SiO2} & 17.657 & $15$ & $239$ & 17.657 & $306$ & $927$ & 17.657 & $7$ & $102$  \\
            CO/MgO (100) & 61.650 & $19$ & $412$ & 61.650 & $187$ & $565$ & 61.650 & $9$ & $180$  \\
            \textit{trans}-$(\ce{C2H2})_{\infty}$ & 3.343 & $4$ & $62$ & 3.343 & $27$ & $80$ & 3.343 & $4$ & $50$  \\
            C-nanotube & 28.141 & $7$ & $150$ & 28.141 & $65$ & $194$ & 28.142 & $8$ & $154$  \\
            Graphene & 2.801 & $5$ & $123$ & 2.801 & $129$ & $386$ & 2.801 & $9$ & $172$  \\
            Aluminum & 2.683 & $12$ & $313$ & 2.684 & $197$ & $592$ & 2.683 & $11$ & $202$  \\
            \bottomrule
        \end{tabular}
    \end{table*}

    \section{Results and discussion}
    \label{sec:results_and_discussion}

    \subsection{Convergence of $k$-CIAH-based PMWF optimization}
    \label{subsec:res_convergence}

    \Cref{tab:conv} compares the convergence behavior of $k$-CIAH and $k$-BFGS for the ten systems listed in \cref{tab:info}.
    Both algorithms converge to the same set of PMWFs, as evidenced by the agreement in the final PM objective values.
    In all cases, $k$-CIAH reaches a stable solution directly, whereas $k$-BFGS initially converges to an unstable solution for MgO, which is subsequently resolved by the stability analysis in \cref{subsec:stability_analysis}.

\begin{figure}[!h]
        \centering
        \includegraphics[width=2.7in]{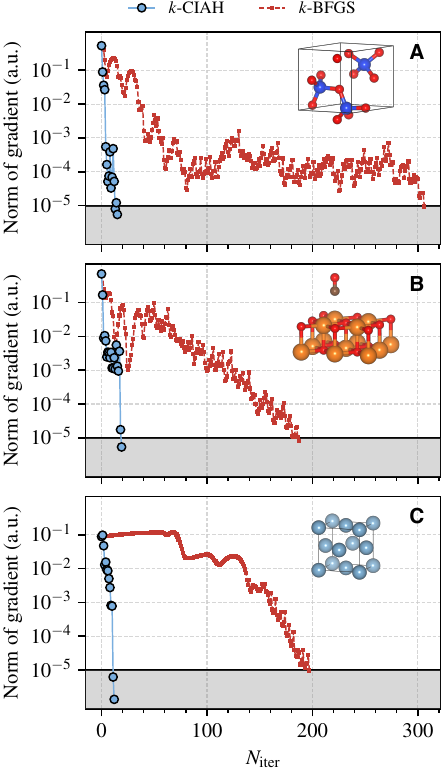}
        \caption{Convergence of $k$-CIAH and $k$-BFGS measured by the decay of the norm of gradient for (A) \ce{SiO2}, (B) CO/MgO(001), and (C) aluminum.
        The convergence threshold ($10^{-5}$) is denoted by the black horizontal line.
        }
        \label{fig:conv}
    \end{figure}

    The number of unitary updates ($N_{\textrm{iter}}$) required for convergence differs markedly between the two methods: the second-order $k$-CIAH algorithm typically converges in $5$--$20$ iterations regardless of system type, while the first-order $k$-BFGS requires $30$--$300$ iterations, roughly an order of magnitude more.
    The trend observed for $k$-BFGS here is consistent with previous studies employing BFGS for PMWF optimization~\cite{Clement21JCTC,Zhu24JPCA}.
    In \cref{fig:conv}, we illustrate for three challenging systems—\ce{SiO2}, CO/MgO(001), and aluminum—that the rapid convergence of $k$-CIAH stems from its ability to take large steps that drive a fast decay of the gradient norm.
    By contrast, the gradient norm in $k$-BFGS fluctuates and only exhibits superlinear convergence close to convergence, as also shown in \cref{fig:conv}.
    \myred{Similar convergence behaviors are observed in the PM objective value for these systems shown in Fig.~S1.}

    \myred{\Cref{tab:conv} further compares $k$-CIAH and $k$-BFGS in terms of the total numbers of PM objective function, gradient, and---for $k$-CIAH only---Hessian--vector product evaluations, collectively denoted as $N_{\mathrm{f}+\mathbf{g}+\mathbf{Hv}}$.
    The breakdown of these three counts is provided in Table~S3 and visualized in Fig.~S2 of the Supporting Information.}
    As discussed in \cref{subsec:cost}, all three operations scale as $O(N_k^2 n^3)$ and dominate the overall computational cost, \myred{although the Hessian--vector product has a smaller prefactor than the other two operations (Table~S1).}
    In all cases except MgO, $k$-CIAH reduces the total $N_{\mathrm{f}+\mathbf{g}+\mathbf{Hv}}$ relative to $k$-BFGS, suggesting higher overall computational efficiency.
    For MgO, $k$-CIAH requires more $N_{\mathrm{f}+\mathbf{g}+\mathbf{Hv}}$ evaluations because it effectively performs an internal stability analysis to escape the unstable solution to which $k$-BFGS initially converges (the cost of this additional stability analysis is not reflected in the $N_{\mathrm{f}+\mathbf{g}}$ reported for $k$-BFGS in \cref{tab:conv}).
    As shown in \cref{subsec:res_cost}, this reduction in $N_{\mathrm{f}+\mathbf{g}+\mathbf{Hv}}$ translates into a more favorable CPU time to solution for $k$-CIAH in practice.

    \Cref{tab:conv} also includes results for $\Gamma$-CIAH for comparison.
    For all systems, $\Gamma$-CIAH converges to the same translationally invariant PMWFs as $k$-CIAH, yielding identical PM objective values.
    \myred{The phase alignment of the initial guess discussed in \cref{subsec:init_guess} was found to be essential for stabilizing the $\Gamma$-CIAH optimization.
    As shown in Table~S5, using the default supercell atomic initial guess without phase alignment leads to convergence to local stationary points in several gapped systems and fails to converge for aluminum within $1000$ cycles.
    For most systems, the convergence behavior of $\Gamma$-CIAH is similar to that of $k$-CIAH, as reflected by the comparable values of $N_{\mathrm{iter}}$ and $N_{\mathrm{f}+\mathbf{g}+\mathbf{Hv}}$.}
    Consequently, because the gradient and Hessian--vector product evaluations in $k$-CIAH scale as $O(N_k^2 n^3)$ rather than $O(N_k^3 n^3)$, the overall cost of $k$-CIAH is lower by a factor of $O(N_k)$ relative to $\Gamma$-CIAH.
    This reduction is demonstrated numerically in \cref{subsec:res_cost}.

    \subsection{Cost of $k$-CIAH-based PMWF optimization}
    \label{subsec:res_cost}

    \begin{figure}[!h]
    \centering
    \includegraphics[width=2.7in]{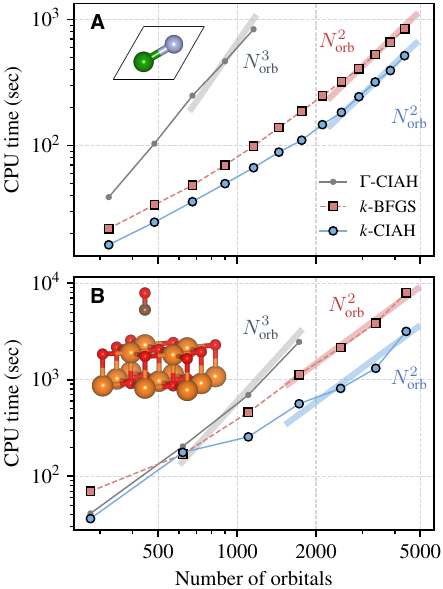}
    \caption{CPU time (seconds) for PMWF optimization using $k$-CIAH, $k$-BFGS, and $\Gamma$-CIAH for (A) h-BN and (B) CO/MgO(001).
    The Brillouin zone is sampled with uniform $n\times n\times 1$ $k$-meshes, with $n=9$--$34$ for h-BN and $n=2$--$8$ for CO/MgO(001).
    Fewer data points are shown for $\Gamma$-CIAH at large $n$ due to memory limitations.
    The $k$-CIAH and $k$-BFGS timings are fitted to an $O(N_{\textrm{orb}}^{2})$ scaling, whereas the $\Gamma$-CIAH data are fitted to an $O(N_{\textrm{orb}}^{3})$ scaling.}
    \label{fig:time}
    \end{figure}

    \Cref{fig:time} compares the CPU time of PMWF optimization using $k$-CIAH, $k$-BFGS, and $\Gamma$-CIAH for two representative systems: h-BN, which has a small unit cell with four occupied bands to be localized, and CO/MgO(001), which has a larger unit cell with 69 occupied bands.
    In both cases, we increase the system size by uniformly refining the $k$-mesh in the $xy$ plane.
    Linear fits in log--log scale confirm the expected asymptotic behavior: $k$-CIAH and $k$-BFGS exhibit quadratic scaling with $N_{\textrm{orb}}$, whereas $\Gamma$-CIAH shows cubic scaling.
    Consistent with the reduction in $N_{\mathrm{f}+\mathbf{g}+\mathbf{Hv}}$ discussed in \cref{subsec:res_cost}, $k$-CIAH is faster than $k$-BFGS by approximately a factor of two to three for both systems.
    For h-BN, both $k$-point methods are substantially faster than $\Gamma$-CIAH due to the small number of occupied bands.
    As the band count increases, however, the performance gap between the $k$-point and $\Gamma$-point approaches narrows: for CO/MgO(001), $\Gamma$-CIAH remains more efficient than $k$-BFGS up to roughly 500 localized orbitals.
    Nevertheless, $k$-CIAH retains a clear advantage over $\Gamma$-CIAH across the range of system sizes considered here.

    \subsection{Wannier interpolation for band structure calculations}
    \label{subsec:res_wannier}

    \begin{figure*}[!t]
    \centering
    \includegraphics[width=5.8in]{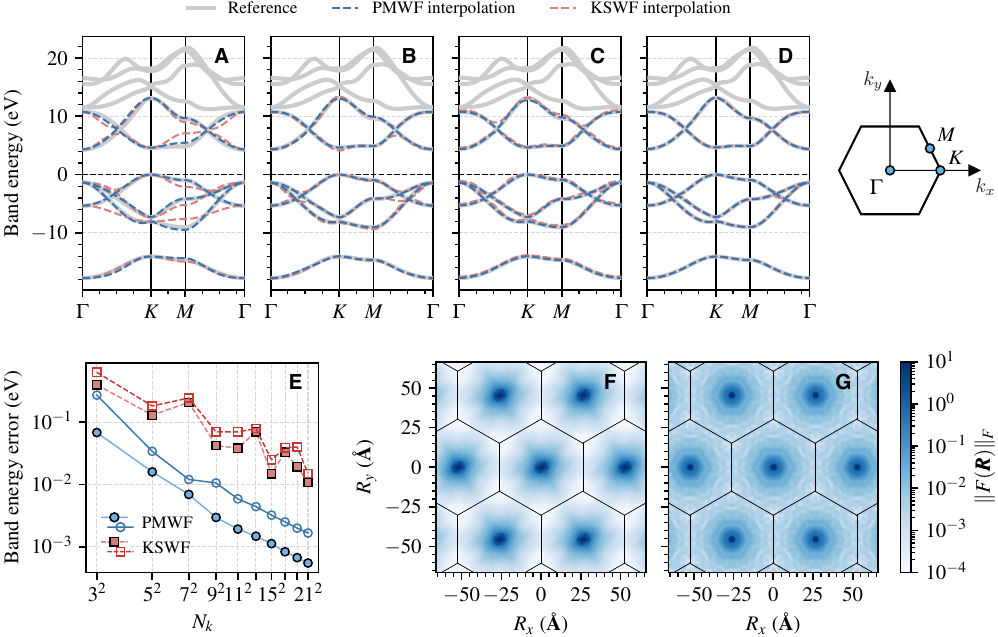}
    \caption{(A--D) Electronic band structure of h-BN obtained from reference non-SCF calculations (gray) and from Wannier interpolation using PMWFs generated by $k$-CIAH (blue) for increasing SCF $k$-mesh sizes: (A) $3\times3$, (B) $5\times5$, (C) $7\times7$, and (D) $9\times9$.
    For comparison, bands interpolated using Kohn--Sham-based WFs without PM localization (KSWFs) are shown in red.
    In all cases, four occupied and two virtual bands are included in the interpolation.
    The high-symmetry points in the first Brillouin zone are indicated on the right.
    (E) Mean absolute error of the interpolated highest-occupied band (HOB; filled markers) and lowest-unoccupied band (LUB; hollow markers) as a function of SCF $k$-mesh size for h-BN.
    (F--G) Frobenius norm of the real-space Fock matrix $\|F(\bm{R})\|_{F}$ in the basis of (F) PMWFs and (G) KSWFs, visualized as two-dimensional heat maps for h-BN using a $21\times21$ SCF $k$-mesh.
    Black lines delineate the boundaries of the Wigner--Seitz cells.}
    \label{fig:band_all}
    \end{figure*}

    \myred{Band interpolation based on maximally localized or Foster--Boys WFs has become a standard practice in the computational physics and materials science literature.}~\cite{Yates07PRB,Marzari12RMP}
    Here, we demonstrate the use of PMWFs for electronic band structure calculations via the Wannier interpolation procedure outlined in \cref{subsec:wannier_interpolation} \myred{using monolayer h-BN as a simple example, for which the four occupied and two lowest-energy virtual bands form a relatively disentangled subspace that facilitates both orbital localization and band interpolation.
    Application to the more general case of entangled bands would in general require a preceding disentanglement step~\cite{Souza01PRB,Marzari12RMP,Damle18MMS}, whose implementation and benchmark are beyond the scope of the present work and are left for future study.}
    \Cref{fig:band_all}(A--D) compares PMWF-interpolated four occupied and two virtual bands with reference bands obtained from non-SCF calculations for h-BN using different SCF $k$-meshes.
    As the SCF $k$-mesh is refined, the PMWF-interpolated bands rapidly approach the reference bands, yielding high-quality band structures even for a relatively coarse $5\times5$ mesh (\cref{fig:band_all}B).
    By contrast, interpolation based on WFs constructed directly from Kohn--Sham orbitals without PM localization \myred{or phase alignment} (hereafter KSWFs) exhibits noticeable deviations from the reference bands especially at band crossing points, even for a $9\times9$ SCF $k$-mesh.
    \myred{We note that aligning the phase of Kohn-Sham orbitals using the procedure described in \cref{subsec:init_guess} does not improve the quality of the interpolated bands.}

    \Cref{fig:band_all}(E) quantifies the band-interpolation error by reporting the mean absolute error of the highest-occupied band (HOB) and lowest-unoccupied band (LUB) as a function of the SCF $k$-mesh size for both PMWF- and KSWF-based interpolation.
    For PMWFs, the error drops well below $0.1$~eV already at a $5\times5$ $k$-mesh and decreases monotonically upon further $k$-mesh refinement.
    In contrast, the error decay for KSWF-based interpolation is more erratic and significantly slower, typically requiring SCF $k$-meshes roughly an order of magnitude denser to reach comparable accuracy.
    This comparison underscores the importance of orbital localization for reliable Wannier interpolation.

    The different interpolation accuracy of PMWFs and KSWFs can be rationalized by examining the spatial decay of the real-space Fock matrix elements defined in \cref{eq:FRij}.
    \Cref{fig:band_all}(F,G) shows $\|F(\bm{R})\|_{F}$ (i.e.,~the Frobenius norm) in both the PMWF and KSWF bases for h-BN with a $21\times21$ SCF $k$-mesh.
    In both representations, the Fock matrix elements decay as one moves from the center toward the boundary of a Wigner--Seitz cell.
    However, the decay is substantially faster in the PMWF basis than in the KSWF basis, which explains the superior band-interpolation performance of PMWFs.

    \section{Conclusion}
    \label{sec:conclusion}

    In summary, we have developed a $k$-point extension of the second-order co-iterative augmented Hessian algorithm, termed $k$-CIAH, for Pipek--Mezey localization of Wannier functions.
    Through an efficient evaluation of the Hessian--vector product, $k$-CIAH achieves $O(N_k^2 n^3)$ scaling in both CPU time and memory, matching that of first-order approaches reported previously~\cite{Clement21JCTC,Zhu24JPCA} while improving upon the $O(N_k^3 n^3)$ scaling of $\Gamma$-point CIAH~\cite{Sun16arXiv}.
    Benchmark calculations on a diverse set of solids demonstrate the fast and robust convergence of $k$-CIAH, making its overall computational efficiency competitive with that of $k$-BFGS and $\Gamma$-CIAH.
    The quality of the resulting PMWFs is further validated by accurate electronic band structures obtained via PMWF-based Wannier interpolation.

    Several avenues for future work remain.
    First, the current $O(N_k^2)$ dependence on the $k$-mesh size can, in principle, be reduced to $O(N_k)$ by adopting a real-space formulation that explicitly exploits the locality of WFs~\cite{Weng22JCTC}.
    This reduction would benefit both $k$-CIAH and $k$-BFGS and enable applications to substantially larger systems than those considered here.
    Second, orbital localization underpins many reduced-scaling correlated wavefunction methods~\cite{Nejad25JCP,Zhu25JCP,Zhu25arXiv,Ye23arXiv,Ye24FD,Ye24JCTC,Hansen21MP,Lau21JPCL,Huang25NCommun}, whose application to periodic solids has expanded rapidly over the past decade.
    High-quality localized WFs may therefore have a significant impact on local correlation treatments of challenging systems, such as metals, bulk defects, and solid interfaces.
    Finally, the $k$-CIAH framework developed in this work can be extended to other periodic calculations that involve orbital optimization, most notably second-order SCF methods~\cite{Werner85JCP}.

    \section*{Supporting Information}

    See the Supporting Information for (i) structural files; (ii) CPU timings for key steps in $k$-point and $\Gamma$-point PMWF optimization; (iii) convergence benchmarks for $k$-BFGS; (iv) a breakdown of the PM objective-function, gradient, and Hessian--vector product evaluations reported in \cref{tab:conv}; (v) convergence benchmarks using alternative atomic projectors; (vi) the effect of the initial guess on the performance of $\Gamma$-CIAH; (vii) convergence of the PM objective value for $k$-CIAH and $k$-BFGS on the three systems in \cref{fig:conv}; (viii) a visualization of the breakdown of PM objective-function, gradient, and Hessian--vector product evaluations reported in \cref{tab:conv}; (ix) derivations of the analytical gradient, Hessian--vector product, and Hessian diagonals of the PM objective function; (x) optimization of the molecular Hessian--vector product evaluation; and (xi) derivation of the Jacobi-sweep-based stability analysis.


    \section*{Conflict of interest}
    The authors declare no competing conflicts of interest.

    \section*{Data availability}

    The data that support the findings of this study are available from the corresponding author upon reasonable request.

    \section*{Acknowledgments}

    This work was supported by the National Science Foundation under Grant No.~CHE-2543461.
    We are grateful for Timothy C.~Berkelbach and Qiming Sun for helpful discussion.
    We acknowledge computing resources provided by the Division of Information Technology at the University of Maryland, College Park.

    \bibliography{refs}

\end{document}


\title{Supporting Information for Fast Generation of Pipek--Mezey Wannier Functions via the Co-Iterative Augmented Hessian Method}

    \author{Gengzhi Yang}
    \affiliation{Joint Center for Quantum Information and Computer Science, University of Maryland, College Park, Maryland, 20742}
    \affiliation{Department of Mathematics, University of Maryland, College Park, Maryland, 20742}

    \author{Hong-Zhou Ye}
    \email{hzye@umd.edu}
    \affiliation{Department of Chemistry and Biochemistry, University of Maryland, College Park, Maryland, 20742}
    \affiliation{Institute for Physical Science and Technology, University of Maryland, College Park, Maryland, 20742}

    \date{\today}

    \maketitle

    \tableofcontents

    \clearpage

    \section{Structure files}

    The structure files for all solid-state systems studied in this work are available in the following GitHub repository:
    \begin{center}
        \href{https://github.com/hongzhouye/supporting\_data/tree/main/2026/PMWF}{https://github.com/hongzhouye/supporting\_data/tree/main/2026/PMWF}
    \end{center}
    where
    \begin{itemize}
        \item the structure of h-BN, diamond, MgO, silicon, \ce{SiO2}, \textit{trans}-$(\ce{C2H2})_{\infty}$, C-nanotube, and graphene is taken from \href{https://doi.org/10.1021/acs.jpca.4c04555}{J.~Phys.~Chem.~A \textbf{128}, 8570 (2024)},

        \item the structure of CO/MgO(001) is taken from \href{https://doi.org/10.1039/D4FD00041B}{Faraday Discuss.~\textbf{254}, 628 (2024)},

        \item and the structure of aluminum (\textit{fcc}) is generated with a lattice constant of $4.05~\textrm{\AA}$.
    \end{itemize}

    \clearpage

    \section{Supplementary tables}

    \begin{table*}[!h]
        \centering
        \caption{CPU times (s) of key steps in $k$-point and $\Gamma$-point PMWF optimization for all systems considered in this work.
        \textbf{Proj.}: atomic projector initialization (for the meta-L{\"o}wdin projectors).
        \textbf{Obj.}: PM objective function evaluation.
        \textbf{Grad.}: gradient evaluation.
        \textbf{Hvp (init)}: precomputation for the Hessian--vector product.
        \textbf{Hvp}: Hessian--vector product evaluation.
        Note that atomic projector initialization is performed only once for the entire optimization, whereas the precomputation for the Hessian--vector product is performed once per CIAH iteration.
        The reported CPU times are averaged over $5$ independent runs using the same computational resources described in Section \fakeref{III} of the main text.\\
        Main conclusion: (i) for the $k$-point based formulation, each Hessian--vector product evaluation is computationally less expensive than that of objective function and gradient evaluation, and (ii) all operations in the $k$-point based formulation are more efficient than their counterparts in the $\Gamma$-point formulation.
        }
        \begin{ruledtabular}
        \begin{tabular}{lrrrrrrrrrr}
            \multirow{2}{*}{System}
            & \multicolumn{5}{c}{$k$-point}
            & \multicolumn{5}{c}{$\Gamma$-point}    \\
            \cmidrule(lr){2-6}
            \cmidrule(lr){7-11}
            & Proj. & Obj. & Grad & Hvp (init) & Hvp
            & Proj. & Obj. & Grad & Hvp (init) & Hvp \\
            \midrule
            h-BN & 29.22 & 0.48 & 0.79 & 0.79 & 0.31 & 154.21 & 2.19 & 5.23 & 6.17 & 3.93  \\
            Diamond & 24.10 & 1.30 & 1.91 & 1.95 & 0.69 & 456.73 & 6.42 & 16.65 & 19.37 & 12.75  \\
            MgO & 43.84 & 0.96 & 1.39 & 1.41 & 0.43 & 749.51 & 8.03 & 17.97 & 20.61 & 12.62  \\
            Silicon & 23.30 & 0.77 & 1.21 & 1.23 & 0.41 & 500.65 & 6.82 & 19.72 & 22.89 & 15.93  \\
            \ce{SiO2} & 4.24 & 0.18 & 0.33 & 0.38 & 0.17 & 28.19 & 0.47 & 1.77 & 1.75 & 1.25  \\
            CO/MgO (100) & 3.54 & 0.18 & 0.41 & 0.47 & 0.28 & 12.88 & 0.30 & 1.00 & 1.36 & 1.01  \\
            \textit{trans}-$(\ce{C2H2})_{\infty}$ & 12.93 & 0.30 & 0.44 & 0.44 & 0.20 & 43.82 & 0.54 & 2.28 & 2.70 & 2.12  \\
            C-nanotube & 7.10 & 0.46 & 0.95 & 1.03 & 0.53 & 79.86 & 1.28 & 3.01 & 3.36 & 2.54  \\
            Graphene & 14.89 & 0.41 & 0.72 & 0.76 & 0.29 & 149.41 & 2.56 & 7.99 & 7.15 & 5.52  \\
            Aluminum & 9.27 & 0.53 & 0.73 & 0.79 & 0.34 & 160.62 & 2.72 & 7.60 & 9.14 & 6.50  \\
        \end{tabular}
        \end{ruledtabular}
    \end{table*}

    \clearpage

    \begin{table*}[!h]
        \centering
        \caption{Number of iterations required for $k$-BFGS to converge as a function of the maximum step size $s_0$, defined in Eqn~\fakeref{38} of the main text.
        In most cases, the converged PMWFs are the same as those reported in Table~\fakeref{3} of the main text.
        Two exceptions are the C-nanotube for $s_0 \geq 0.30$~a.u.~and aluminum for $s_0 = 0.20$~a.u.,~where local minima with a higher PM objective value are obtained.
        To balance efficiency and numerical stability, we adopt $s_0 = 0.10$~a.u. as the default.
        }
        \begin{ruledtabular}
        \begin{tabular}{lrrrrrrrrrr}
            \multirow{2}{*}{System}
            & \multicolumn{6}{c}{Maximum step size, $s_0$ (a.u.)}    \\
            \cmidrule(lr){2-7}
            & $0.05$ & $0.10$ & $0.20$ & $0.30$ & $0.50$ & $1.00$   \\
            \midrule
            h-BN & 38 & 27 & 22 & 19 & 20 & 20  \\
            Diamond & 70 & 51 & 34 & 27 & 21 & 119  \\
            MgO & 32 & 24 & 23 & 22 & 22 & 22  \\
            Silicon & 65 & 49 & 32 & 25 & 314 & 160  \\
            \ce{SiO2} & 374 & 312 & 425 & 350 & 259 & 303  \\
            CO/MgO (100) & 287 & 187 & 204 & 212 & 200 & 208  \\
            \textit{trans}-$(\ce{C2H2})_{\infty}$ & 39 & 27 & 35 & 19 & 31 & 17  \\
            C-nanotube & 93 & 65 & 55 & 59$^*$ & 50$^*$ & 49$^*$  \\
            Graphene & 139 & 129 & 123 & 118 & 118 & 117  \\
            Aluminum & 336 & 197 & 198$^*$ & 181 & 182 & 240  \\
        \end{tabular}
        \end{ruledtabular}
    \end{table*}

    \clearpage

    \begin{table*}[!h]
        \centering
        \caption{Same as Table~\fakeref{3} in the main text, except that the breakdown of PM objective function, gradient, and Hessian--vector product evaluations is shown.
        }
        \begin{ruledtabular}
        \begin{tabular}{llllllllll}
            \multirow{2}{*}{System}
            & \multicolumn{3}{c}{$k$-CIAH}
            & \multicolumn{3}{c}{$k$-BFGS}
            & \multicolumn{3}{c}{$\Gamma$-CIAH} \\
            \cmidrule(lr){2-4}
            \cmidrule(lr){5-7}
            \cmidrule(lr){8-10}
            & PM obj. & $N_{\textrm{iter}}$ & $N_{\mathrm{f}}+N_{\mathbf{g}}+N_{\mathbf{Hv}}$
            & PM obj. & $N_{\textrm{iter}}$ & $N_{\mathrm{f}}+N_{\mathbf{g}}$
            & PM obj. & $N_{\textrm{iter}}$ & $N_{\mathrm{f}}+N_{\mathbf{g}}+N_{\mathbf{Hv}}$  \\
            \midrule
            h-BN & 2.276 & $3$ & $3+11+42$ & 2.276 & $27$ & $53+27$ & 2.276 & $6$ & $6+14+50$  \\
            Diamond & 1.859 & $4$ & $4+15+62$ & 1.859 & $51$ & $101+51$ & 1.859 & $5$ & $5+15+57$  \\
            MgO & 3.241 & $9$ & $9+24+105$ & 3.241 & $19+5$ & $46+24$ & 3.241 & $9$ & $9+26+80$  \\
            Silicon & 1.798 & $5$ & $5+18+66$ & 1.798 & $49$ & $97+49$ & 1.798 & $5$ & $5+16+60$  \\
            \ce{SiO2} & 17.657 & $15$ & $15+40+184$ & 17.657 & $306$ & $621+306$ & 17.657 & $7$ & $7+21+74$  \\
            CO/MgO (100) & 61.650 & $19$ & $19+68+325$ & 61.650 & $187$ & $378+187$ & 61.650 & $9$ & $9+34+137$  \\
            \textit{trans}-$(\ce{C2H2})_{\infty}$ & 3.343 & $4$ & $4+13+45$ & 3.343 & $27$ & $53+27$ & 3.343 & $4$ & $4+11+35$  \\
            C-nanotube & 28.141 & $7$ & $7+26+117$ & 28.141 & $65$ & $129+65$ & 28.142 & $8$ & $8+30+116$  \\
            Graphene & 2.801 & $5$ & $5+22+96$ & 2.801 & $129$ & $257+129$ & 2.801 & $9$ & $9+30+133$  \\
            Aluminum & 2.683 & $12$ & $12+52+249$ & 2.684 & $197$ & $395+197$ & 2.683 & $11$ & $11+37+154$  \\
        \end{tabular}
        \end{ruledtabular}
    \end{table*}

    \clearpage

    \begin{table*}[!h]
        \centering
        \caption{Same as Table~\fakeref{3} in the main text, except that intrinsic atomic orbitals are used as the atomic projectors.
        The relative convergence behavior of the three algorithms considered, $k$-CIAH, $k$-BFGS, and $\Gamma$-CIAH, closely follows the pattern observed in Table~\fakeref{3} of the main text, indicating that PMWF optimization is relatively insensitive to the choice of local projectors (here, IAOs versus meta-L{\"o}wdin projectors).
        }
        \begin{ruledtabular}
        \begin{tabular}{llllllllll}
            \multirow{2}{*}{System}
            & \multicolumn{3}{c}{$k$-CIAH}
            & \multicolumn{3}{c}{$k$-BFGS}
            & \multicolumn{3}{c}{$\Gamma$-CIAH} \\
            \cmidrule(lr){2-4}
            \cmidrule(lr){5-7}
            \cmidrule(lr){8-10}
            & PM obj. & $N_{\textrm{iter}}$ & $N_{\mathrm{f}}+N_{\mathbf{g}}+N_{\mathbf{Hv}}$
            & PM obj. & $N_{\textrm{iter}}$ & $N_{\mathrm{f}}+N_{\mathbf{g}}$
            & PM obj. & $N_{\textrm{iter}}$ & $N_{\mathrm{f}}+N_{\mathbf{g}}+N_{\mathbf{Hv}}$  \\
            \midrule
            h-BN & 2.130 & $4$ & $4+18+67$ & 2.130 & $27$ & $54+27$ & 2.130 & $5$ & $5+14+52$  \\
            Diamond & 1.910 & $5$ & $5+23+89$ & 1.910 & $67$ & $133+67$ & 1.910 & $5$ & $5+14+56$  \\
            MgO & 2.680 & $6$ & $6+30+118$ & 2.680 & $99$ & $205+99$ & 2.680 & $6$ & $6+20+70$  \\
            Silicon & 1.876 & $4$ & $4+20+88$ & 1.876 & $41$ & $81+41$ & 1.876 & $6$ & $6+16+56$  \\
            \ce{SiO2} & 17.849 & $15$ & $15+48+229$ & 17.849 & $251$ & $517+251$ & 17.849 & $8$ & $8+24+85$  \\
            CO/MgO (100) & 57.368 & $13$ & $13+42+214$ & 57.368 & $359$ & $735+359$ & 57.365 & $8$ & $8+28+110$  \\
            \textit{trans}-$(\ce{C2H2})_{\infty}$ & 3.578 & $4$ & $4+17+66$ & 3.578 & $28$ & $56+28$ & 3.578 & $4$ & $4+10+33$  \\
            C-nanotube & 28.576 & $7$ & $7+32+142$ & 28.575 & $66$ & $132+66$ & 28.576 & $7$ & $7+23+90$  \\
            Graphene & 2.907 & $5$ & $5+27+127$ & 2.907 & $123$ & $248+123$ & 2.907 & $15$ & $15+46+189$  \\
            Aluminum & 2.185 & $14$ & $14+64+296$ & 2.190 & $197$ & $394+197$ & 2.190 & $23$ & $23+64+266$  \\
        \end{tabular}
        \end{ruledtabular}
    \end{table*}

    \begin{table*}[!h]
        \centering
        \caption{Same as Table~\fakeref{3} in the main text, except that the convergence behavior of $\Gamma$-CIAH is compared using atomic initial guesses with and without phase alignment (PA).
        The ``with PA'' results are identical to those reported in Table~\fakeref{3}.
        The ``without PA'' results are obtained from the atomic initial guess generated directly in the supercell, which is the default initial guess when the molecular CIAH code is applied directly to a periodic supercell with $\Gamma$-point Brillouin-zone sampling.
        The data suggest that the phase-aligned initial guess stabilizes the $\Gamma$-CIAH optimization by helping avoid convergence to local minima or saddle points (e.g.,~h-BN, diamond, and MgO).
        In the aluminum case, phase alignment enables successful convergence of an otherwise failed optimization after $1000$ cycles.
        }
        \begin{ruledtabular}
        \begin{tabular}{lllllll}
            \multirow{2}{*}{System}
            & \multicolumn{3}{c}{$\Gamma$-CIAH (with PA)}
            & \multicolumn{3}{c}{$\Gamma$-CIAH (without PA)} \\
            \cmidrule(lr){2-4}
            \cmidrule(lr){5-7}
            & PM obj. & $N_{\textrm{iter}}$ & $N_{\mathrm{f}}+N_{\mathbf{g}}+N_{\mathbf{Hv}}$
            & PM obj. & $N_{\textrm{iter}}$ & $N_{\mathrm{f}}+N_{\mathbf{g}}+N_{\mathbf{Hv}}$   \\
            \midrule
            h-BN & 2.276 & $6$ & $6+14+50$ & 2.276 & $4+3$ & $7+14+38$  \\
            Diamond & 1.859 & $5$ & $5+15+57$ & 1.859 & $10+7$ & $17+36+140$  \\
            MgO & 3.241 & $9$ & $9+26+80$ & 3.241 & $4+3$ & $7+15+45$  \\
            Silicon & 1.798 & $5$ & $5+16+60$ & 1.798 & $7$ & $7+17+60$  \\
            \ce{SiO2} & 17.657 & $7$ & $7+21+74$ & 17.657 & $7$ & $7+20+73$  \\
            CO/MgO (100) & 61.650 & $9$ & $9+34+137$ & 61.650 & $7+3$ & $10+27+92$  \\
            \textit{trans}-$(\ce{C2H2})_{\infty}$ & 3.343 & $4$ & $4+11+35$ & 3.343 & $5$ & $5+11+29$  \\
            C-nanotube & 28.142 & $8$ & $8+30+116$ & 28.163 & $7$ & $7+21+88$  \\
            Graphene & 2.801 & $9$ & $9+30+133$ & 2.800 & $22$ & $22+72+342$  \\
            Aluminum & 2.683 & $11$ & $11+37+154$ & \multicolumn{3}{c}{Failed}  \\
        \end{tabular}
        \end{ruledtabular}
    \end{table*}

    \clearpage

    \section{Supplementary figures}

    \begin{figure}[!h]
        \centering
        \includegraphics[width=3.3in]{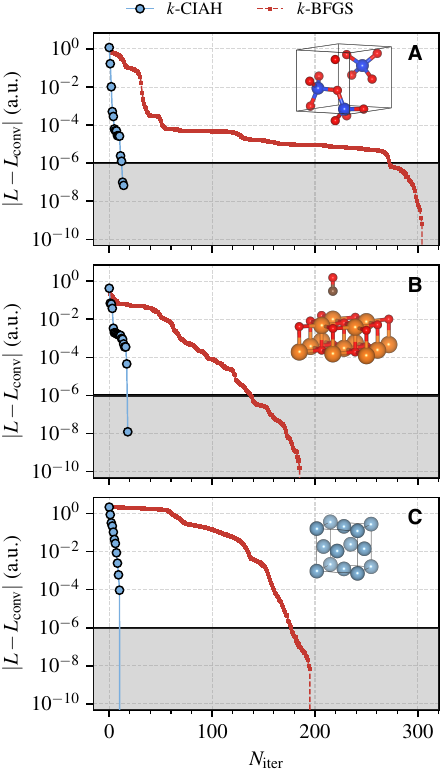}
        \caption{Convergence of $k$-CIAH and $k$-BFGS measured by the decay of the absolute error of PM objective function $L$ for (A) \ce{SiO2}, (B) CO/MgO(001), and (C) aluminum.
        The convergence threshold ($10^{-6}$) is denoted by the black horizontal line.}
    \end{figure}

    \begin{figure}[!h]
        \centering
        \includegraphics[width=1.0\linewidth]{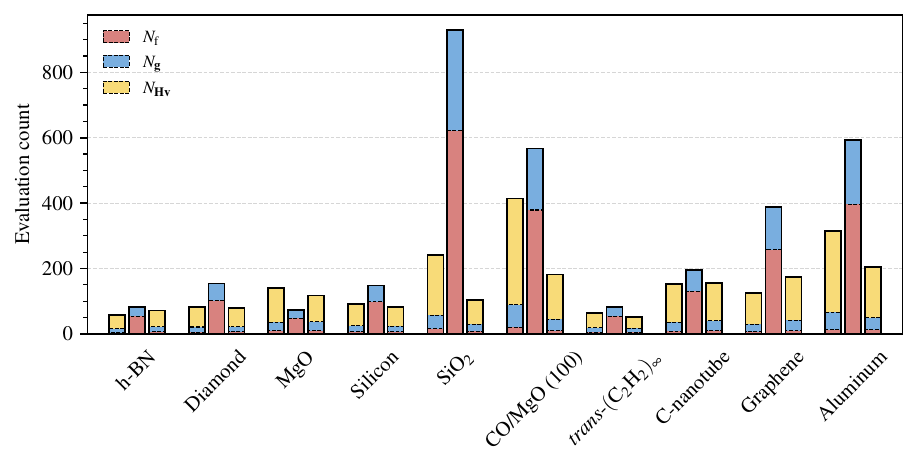}
        \caption{Breakdown of the evaluation count of PM objective ($N_{\mathrm{f}}$), gradient ($N_{\mathbf{g}}$), and Hessian--vector product ($N_{\mathrm{Hv}}$) for the ten systems discussed in Table~\fakeref{3} in the main text.
        The total count, i.e.,~the sum of the three numbers, for each system is the number reported in Table~\fakeref{3} in the main text.}
    \end{figure}

    \clearpage

    \section{Derivation of analytical gradients, Hessian--vector product, and Hessian diagonals}

    \subsection{Parameterization}

    As discussed in the main text, the $k$-space unitaries can be parameterized using $k$-dependent generators, each of which is a complex anti-Hermitian matrix.
    Let $\kappa_{\bm{k}} = X_{\bm{k}} + \mi~Y_{\bm{k}}$ where both $X_{\bm{k}}$ and $Y_{\bm{k}} \in \mathbb{R}^{n_{\textrm{orb}} \times n_{\textrm{orb}}}$.
    The unique parameters are the lower-triangular part of $X_{\bm{k}}$, excluding the diagonals, and the the lower-triangular part of $Y_{\bm{k}}$, including the diagonals.

    \subsection{Basic derivatives}

    Define $(E_{mn})_{ij} = \delta_{mi}\delta_{nj}$.
    The first- and second-order derivatives of the unitary with respect to the generators evaluated at $\kappa_{\bm{k}} = 0$ are
    \begin{equation}    \label{eq:U_first_deriv}
    \begin{split}
        \frac{\partial U_{\bm{k}'}}{\partial X_{\bm{k}mn}}
            &= \delta_{\bm{k}\bm{k}'} (E_{mn} - E_{nm}),    \qquad{} (m \neq n) \\
        \frac{\partial U_{\bm{k}'}}{\partial Y_{\bm{k}mn}}
            &= \mi \delta_{\bm{k}\bm{k}'} (E_{mn} + E_{nm}),  \qquad{} (m \neq n)  \\
        \frac{\partial U_{\bm{k}'}}{\partial Y_{\bm{k}mm}}
            &= \mi \delta_{\bm{k}\bm{k}'} E_{mm}.\\
    \end{split}
    \end{equation}
    \begin{equation}    \label{eq:U_second_deriv}
    \begin{split}
        \frac{\partial^2 U_{\bm{k}''}}{\partial A_{\bm{k}mn} B_{\bm{k}'pq}}
            = \frac{1}{2} \left\{
                \frac{\partial U_{\bm{k}''}}{\partial A_{\bm{k}mn}},
                \frac{\partial U_{\bm{k}''}}{\partial B_{\bm{k}'mn}}
            \right\},
        \qquad{}(A, B = X, Y)
    \end{split}
    \end{equation}

    \subsection{Cost function}

    The PM objective function defined in the main text here is recapped here.
    A minus sign is added to the objective function to turn the maximization a minimization problem.
    \begin{equation}
        L
            = -\sum_{\bm{T}A,i} Q_{\bm{T}A,\bm{0}i}^{p}
    \end{equation}
    \begin{equation}
        Q_{\bm{T}A,\bm{0}i}
            = P_{\bm{T}A,\bm{0}i,\bm{0}i}
            = \sum_{\bm{k}\bm{k}'} \left(
                U^{\dagger}_{\bm{k}} P_{\bm{T}A,\bm{k}\bm{k}'} U_{\bm{k}'}
            \right)_{ii}
    \end{equation}
    \begin{equation}
        P_{\bm{T}A,\bm{k}\bm{k}'}
            = \frac{1}{N_k} \sum_{\bm{R}\bm{R}'} \theta_{\bm{R}\bm{k}}^* P_{\bm{T}A,\bm{R}\bm{R}'} \theta_{\bm{R}'\bm{k}'}
    \end{equation}

    \subsection{Gradient}

    Consider the first-order variation of $L$,
    \begin{equation}
        \delta L
            = -p\sum_{\bm{T}A,i} Q_{\bm{T}A,\bm{0}i}^{p-1} \delta Q_{\bm{T}A,\bm{0}i}
    \end{equation}
    which depends on the first-order variation of $Q$,
    \begin{equation}
        \delta Q_{\bm{T}A,\bm{0}i}
            = 2\Re \sum_{\bm{k}'} \left(
                P_{\bm{T}A,\bm{0}\bm{k}'} \delta U_{\bm{k}'}
            \right)_{ii},
        \qquad{}\left(
            P_{\bm{T}A,\bm{0}\bm{k}'}
                := \sum_{\bm{k}} P_{\bm{T}A,\bm{k}\bm{k}'}
        \right)
    \end{equation}
    For the real and imaginary part of the generator, the derivatives are
    \begin{equation}    \label{eq:Q_grad}
    \begin{split}
        \frac{\partial Q_{\bm{T}A,\bm{0}i}}{\partial X_{\bm{k}mn}}
            &= -2 \left(
                P^{\Re}_{\bm{T}A,\bm{0}m,\bm{k}n} \delta_{mi} -
                P^{\Re}_{\bm{T}A,\bm{0}n,\bm{k}m} \delta_{ni}
            \right) \\
        \frac{\partial Q_{\bm{T}A,\bm{0}i}}{\partial Y_{\bm{k}mn}}
            &= -2 \left(
                P^{\Im}_{\bm{T}A,\bm{0}m,\bm{k}n} \delta_{mi}
                + P^{\Im}_{\bm{T}A,\bm{0}n,\bm{k}m} \delta_{ni}
            \right)
    \end{split}
    \end{equation}
    The two can be combined conveniently into a complex-valued expression,
    \begin{equation}
        \frac{\partial Q_{\bm{T}A,\bm{0}i}}{\partial X_{\bm{k}mn}} + \mi \frac{\partial Q_{\bm{T}A,\bm{0}i}}{\partial Y_{\bm{k}mn}}
            = -2 \left(
                P_{\bm{T}A,\bm{0}m,\bm{k}n} \delta_{mi}
                - P_{\bm{T}A,\bm{0}n,\bm{k}m}^* \delta_{ni}
            \right)
    \end{equation}
    which leads to the following expression for the gradient,
    \begin{equation}    \label{eq:gradient_final}
    \boxed{
    \begin{split}
        G_{\bm{k}}
            &= \tilde{G}_{\bm{k}} - \tilde{G}_{\bm{k}}^{\dagger},   \\
        \tilde{G}_{\bm{k}mn}
            &= 2p \sum_{\bm{T}A} Q_{\bm{T}A,\bm{0}m}^{p-1} P_{\bm{T}A,\bm{0}m,\bm{k}n}
    \end{split}
    }
    \end{equation}
    Using $P_{\bm{T}A,\bm{k}\bm{0}} = P_{\bm{T}A,\bm{0}\bm{k}}^{\dagger}$, one can convert \cref{eq:gradient_final} to Eqn~\fakeref{25} in the main text.
    We prefer to put $\bm{k}$ index first as in $P_{\bm{T}A,\bm{k}\bm{0}}$ due to the C-ordering of matrices and tensors in our code.

    \subsection{Hessian--vector product}

    Let $v_{\bm{k}}$ denote the vector to be applied by the Hessian.
    Let $v^{X}_{\bm{k}}$ and $v^{Y}_{\bm{k}}$ denote its real and imaginary parts, respectively.
    The Hessian--vector product is defined as follows.
    \begin{equation}
    \begin{split}
        \Sigma^X_{\bm{k}mn}
            &= \frac{1}{2} \sum_{\bm{k}'pq} \frac{\partial^2 L}{\partial X_{\bm{k}mn} \partial X_{\bm{k}'pq}} v^{X}_{\bm{k}'pq} +
            \frac{\partial^2 L}{\partial X_{\bm{k}mn} \partial Y_{\bm{k}'pq}}  v^{Y}_{\bm{k}'pq},    \\
        \Sigma^Y_{\bm{k}mn}
            &= \frac{1}{2} \sum_{\bm{k}'pq} \frac{\partial^2 L}{\partial Y_{\bm{k}mn} \partial X_{\bm{k}'pq}}  v^{X}_{\bm{k}'pq} +
            \frac{\partial^2 L}{\partial Y_{\bm{k}mn} \partial Y_{\bm{k}'pq}}  v^{Y}_{\bm{k}'pq},    \\
    \end{split}
    \end{equation}
    where at a high-level, we need to evaluate the second-order variation of $L$,
    \begin{equation}
        \delta^2 L
            = -p(p-1) \sum_{\bm{T}Ai} Q_{\bm{T}A,\bm{0}i}^{p-2} \left(\delta Q_{\bm{T}A,\bm{0}i}\right)^2
            - p \sum_{\bm{T}Ai} Q_{\bm{T}A,\bm{0}i}^{p-1} \delta^2 Q_{\bm{T}A,\bm{0}i}
    \end{equation}
    The first term only depends on the first-order variation of $Q$, whose expression has already been derived above.
    We call this term \textit{disconnected}.
    The second term depends on the second-order variation of $Q$,
    \begin{equation}
        \delta^2 Q_{\bm{T}A,\bm{0}i}
            = 2 \Re \sum_{\bm{k}'} \left(
                P_{\bm{T}A,\bm{0}\bm{k}'} \delta^2 U_{\bm{k}'}
            \right)_{ii}
            + 2 \Re \sum_{\bm{k}\bm{k}'} \left(
                \delta U^{\dagger}_{\bm{k}} P_{\bm{T}A,\bm{k}\bm{k}'} \delta U_{\bm{k}'}
            \right)_{ii}
    \end{equation}
    and therefore is \textit{connected}.
    We further differentiate the first and second term in $\delta^2 Q$ to be \textit{asymmetric} and \textit{symmetric} connected terms.

    \subsubsection{Disconnected term}

    \begin{equation}
    \begin{split}
        \Sigma^{X}_{\bm{k}mn}
            &= -\frac{1}{2} p(p-1) \sum_{\bm{T}Ai} Q_{\bm{T}A,\bm{0}i}^{p-2} \frac{\partial Q_{\bm{T}A,\bm{0}i}}{\partial X_{\bm{k}mn}}
            \sum_{\bm{k}'pq} \left(
                \frac{\partial Q_{\bm{T}A,\bm{0}i}}{\partial X_{\bm{k}'pq}} v^{X}_{\bm{k}'pq}
                + \frac{\partial Q_{\bm{T}A,\bm{0}i}}{\partial Y_{\bm{k}'pq}} v^{Y}_{\bm{k}'pq}
            \right),    \\
        \Sigma^{Y}_{mn}
            &= -\frac{1}{2} p(p-1) \sum_{\bm{T}Ai} Q_{\bm{T}A,\bm{0}i}^{p-2} \frac{\partial Q_{\bm{T}A,\bm{0}i}}{\partial Y_{\bm{k}mn}}
            \sum_{\bm{k}'pq} \left(
                \frac{\partial Q_{\bm{T}A,\bm{0}i}}{\partial X_{\bm{k}'pq}} v^{X}_{\bm{k}'pq}
                + \frac{\partial Q_{\bm{T}A,\bm{0}i}}{\partial Y_{\bm{k}'pq}} v^{Y}_{\bm{k}'pq}
            \right)
    \end{split}
    \end{equation}
    where the term in the parenthesis can be simplified to
    \begin{equation}
        \sum_{\bm{k}'pq} \left(
            \frac{\partial Q_{\bm{T}A,\bm{0}i}}{\partial X_{\bm{k}'pq}} v^{X}_{\bm{k}'pq}
            + \frac{\partial Q_{\bm{T}A,\bm{0}i}}{\partial Y_{\bm{k}'pq}} v^{Y}_{\bm{k}'pq}
        \right)
            = 4 \Re \sum_{\bm{k}'} \left(
                P_{\bm{T}A,\bm{0}\bm{k}'} v_{\bm{k}'}
            \right)_{ii}
    \end{equation}
    Combining this with the gradient of $Q$ [\cref{eq:Q_grad}] gives
    \begin{equation}    \label{eq:Hvp_disconn_final}
    \boxed{
    \begin{split}
        \Sigma_{\bm{k}}
            &= \tilde{\Sigma}_{\bm{k}} - \tilde{\Sigma}_{\bm{k}}, \\
        \tilde{\Sigma}_{\bm{k}mn}
            &= 4 p(p-1) \sum_{\bm{T}A} Q_{\bm{T}A,\bm{0}m}^{p-2} \sum_{\bm{k}'} \Re(P_{\bm{T}A,\bm{0}\bm{k}'} v_{\bm{k}'})_{mm} P_{\bm{T}A,\bm{0}m,\bm{k}n}
    \end{split}
    }
    \end{equation}

    \subsubsection{Connected symmetric term}

    \begin{equation}    \label{eq:kpt_sigmaXY_raw}
    \begin{split}
        \Sigma^{X}_{\bm{k}mn}
            &= -p \sum_{\bm{T}A,i} Q_{\bm{T}A,\bm{0}i}^{p-1} \Re \left[
                \frac{\partial U_{\bm{k}}^{\dagger}}{\partial X_{\bm{k}mn}} P_{\bm{T}A,\bm{k}\bm{k}'}
                \sum_{\bm{k}'pq} \left(
                    \frac{\partial U_{\bm{k}'}}{\partial X_{\bm{k}'pq}} v^{X}_{\bm{k}'pq}
                    + \frac{\partial U_{\bm{k}'}}{\partial Y_{\bm{k}'pq}} v^{Y}_{\bm{k}'pq}
                \right)
            \right]_{ii},   \\
        \Sigma^{Y}_{\bm{k}mn}
            &= -p \sum_{\bm{T}A,i} Q_{\bm{T}A,\bm{0}i}^{p-1} \Re \left[
                \frac{\partial U_{\bm{k}}^{\dagger}}{\partial Y_{\bm{k}mn}} P_{\bm{T}A,\bm{k}\bm{k}'}
                \sum_{\bm{k}'pq} \left(
                    \frac{\partial U_{\bm{k}'}}{\partial X_{\bm{k}'pq}} v^{X}_{\bm{k}'pq}
                    + \frac{\partial U_{\bm{k}'}}{\partial Y_{\bm{k}'pq}} v^{Y}_{\bm{k}'pq}
                \right)
            \right]_{ii},   \\
    \end{split}
    \end{equation}
    where the term in the parenthesis can be simplified to
    \begin{equation}    \label{eq:kpt_dot_dUdxy_vxy}
        \sum_{\bm{k}'pq} \frac{\partial U_{\bm{k}'}}{\partial X_{\bm{k}'pq}} v^{X}_{\bm{k}'pq} + \frac{\partial U_{\bm{k}'}}{\partial Y_{\bm{k}'pq}} v^{Y}_{\bm{k}'pq}
            = 2 \sum_{\bm{k}'pq} E_{pq} v_{\bm{k}'pq}
    \end{equation}
    Combining this with the gradient of $U$ [\cref{eq:U_first_deriv}] gives
    \begin{equation}    \label{eq:Hvp_conn_symm_final}
    \boxed{
    \begin{split}
        \Sigma_{\bm{k}}
            &= \tilde{\Sigma}_{\bm{k}} - \tilde{\Sigma}_{\bm{k}}^{\dagger}, \\
        \tilde{\Sigma}_{\bm{k}mn}
            &= -2p \sum_{\bm{T}A} Q_{\bm{T}A,\bm{0}n}^{p-1} \sum_{\bm{k}'} \left(P_{\bm{T}A,\bm{k}\bm{k}'} v_{\bm{k}'}\right)_{mn}
    \end{split}
    }
    \end{equation}

    \subsubsection{Connected asymmetric term}

    \begin{equation}
    \begin{split}
        \Sigma^{X}_{\bm{k}mn}
            &= -\frac{1}{2}p \sum_{\bm{T}A,i} Q_{\bm{T}A,\bm{0}i}^{p-1} \Re\left[
                P_{\bm{T}A,\bm{0}\bm{k}} \left\{
                    \frac{\partial U_{\bm{k}}}{\partial X_{\bm{k}mn}},
                    \sum_{pq} \frac{\partial U_{\bm{k}}}{\partial X_{\bm{k}pq}} v^{X}_{\bm{k}pq}
                    + \frac{\partial U_{\bm{k}}}{\partial Y_{\bm{k}pq}} v^{Y}_{\bm{k}pq}
                \right\}
            \right]_{ii}    \\
        \Sigma^{Y}_{\bm{k}mn}
            &= -\frac{1}{2}p \sum_{\bm{T}A,i} Q_{\bm{T}A,\bm{0}i}^{p-1} \Re\left[
                P_{\bm{T}A,\bm{0}\bm{k}} \left\{
                    \frac{\partial U_{\bm{k}}}{\partial Y_{\bm{k}mn}},
                    \sum_{pq} \frac{\partial U_{\bm{k}}}{\partial X_{\bm{k}pq}} v^{X}_{\bm{k}pq}
                    + \frac{\partial U_{\bm{k}}}{\partial Y_{\bm{k}pq}} v^{Y}_{\bm{k}pq}
                \right\}
            \right]_{ii}
    \end{split}
    \end{equation}
    Using \cref{eq:kpt_dot_dUdxy_vxy} to simplify terms in the parenthesis leads to
    \begin{equation}
    \begin{split}
        \Sigma^{X}_{\bm{k}mn}
            &= -p \sum_{\bm{T}A,i} Q_{\bm{T}A,\bm{0}i}^{p-1} \Re\left[
                P_{\bm{T}A,\bm{0}\bm{k}} \left\{
                    \frac{\partial U_{\bm{k}}}{\partial X_{\bm{k}mn}},
                    \sum_{pq} E_{pq} v_{\bm{k}pq}
                \right\}
            \right]_{ii}    \\
        \Sigma^{Y}_{\bm{k}mn}
            &= -p \sum_{\bm{T}A,i} Q_{\bm{T}A,\bm{0}i}^{p-1} \Re\left[
                P_{\bm{T}A,\bm{0}\bm{k}} \left\{
                    \frac{\partial U_{\bm{k}}}{\partial Y_{\bm{k}mn}},
                    \sum_{pq} E_{pq} v_{\bm{k}pq}
                \right\}
            \right]_{ii}    \\
    \end{split}
    \end{equation}
    Using \cref{eq:U_first_deriv} to further simplify the expression leads to
    \begin{equation}    \label{eq:Hvp_conn_asymm_final}
    \boxed{
    \begin{split}
        \Sigma_{\bm{k}}
            &= \tilde{\Sigma}_{\bm{k}} - \tilde{\Sigma}_{\bm{k}}^{\dagger}, \\
        \tilde{\Sigma}_{\bm{k}mn}
            &= p \sum_{i} v_{\bm{k}mi} \left[ \sum_{\bm{T}A} Q_{\bm{T}A,\bm{0}i}^{p-1} (P_{\bm{T}A,\bm{0}\bm{k}})_{in} \right]
            + p \sum_{\bm{T}A} Q_{\bm{T}A,\bm{0}m}^{p-1} (P_{\bm{T}A,\bm{0}\bm{k}} v_{\bm{k}})_{mn}  \\
    \end{split}
    }
    \end{equation}

    Finally, using $P_{\bm{T}A,\bm{k}\bm{0}} = P_{\bm{T}A,\bm{0}\bm{k}}^{\dagger}$, one can convert \cref{eq:Hvp_disconn_final,eq:Hvp_conn_symm_final,eq:Hvp_conn_asymm_final} to Eqn~\fakeref{27--29} in the main text.

    \subsection{Hessian diagonal}

    The Hessian diagonal is defined as:
    \begin{equation}
    \begin{split}
        D^{X}_{\bm{k}mn}
            &= \frac{\partial^2 L}{\partial X_{\bm{k}mn}^2}
            = \tilde{D}_{\bm{k}}^{X} + \tilde{D}_{\bm{k}}^{X\top} \\
        D^{Y}_{\bm{k}mn}
            &= \frac{\partial^2 L}{\partial Y_{\bm{k}mn}^2}
            = \tilde{D}_{\bm{k}}^{Y} + \tilde{D}_{\bm{k}}^{Y\top}
    \end{split}
    \end{equation}
    where the second equality explicitly shows its symmetric nature.

    \subsubsection{Disconnected term}

    \begin{equation}
    \begin{split}
        D^{X}_{\bm{k}mn}
            &= -p(p-1) \sum_{\bm{T}A,i} Q_{\bm{T}A,\bm{0}i}^{p-2} \left(\frac{\partial Q_{\bm{T}A,\bm{0}i}}{\partial X_{\bm{k}mn}}\right)^2 \\
            &= -4p(p-1) \sum_{\bm{T}A,i} Q_{\bm{T}A,\bm{0}i}^{p-2} \left[\Re (P_{\bm{T}A,\bm{0}\bm{k}})_{mn}\right]^2 \delta_{mi} + (m \leftrightarrow n)  \\
            &= -4p(p-1) \sum_{\bm{T}A} Q_{\bm{T}A,\bm{0}m}^{p-2} \left[\Re (P_{\bm{T}A,\bm{0}\bm{k}})_{mn}\right]^2 + (m \leftrightarrow n)  \\
    \end{split}
    \end{equation}
    \begin{equation}
    \begin{split}
        D^{Y}_{\bm{k}mn}
            &= -p(p-1) \sum_{\bm{T}A,i} Q_{\bm{T}A,\bm{0}i}^{p-2} \left(\frac{\partial Q_{\bm{T}A,\bm{0}i}}{\partial Y_{\bm{k}mn}}\right)^2 \\
            &= -4p(p-1) \sum_{\bm{T}A,i} Q_{\bm{T}A,\bm{0}i}^{p-2} \left[\Im (P_{\bm{T}A,\bm{0}\bm{k}})_{mn}\right]^2 \delta_{mi} + (m \leftrightarrow n)  \\
            &= -4p(p-1) \sum_{\bm{T}A} Q_{\bm{T}A,\bm{0}m}^{p-2} \left[\Im (P_{\bm{T}A,\bm{0}\bm{k}})_{mn}\right]^2 + (m \leftrightarrow n)  \\
    \end{split}
    \end{equation}
    \begin{equation}
    \boxed{
    \begin{split}
        \tilde{D}^{X}_{mn}
            &= -4p(p-1) \sum_{\bm{T}A} Q_{\bm{T}A,\bm{0}m}^{p-2} \left[\left(\Re P_{\bm{T}A,\bm{0}\bm{k}}\right)_{mn}\right]^2  \\
        \tilde{D}^{Y}_{mn}
            &= -4p(p-1) \sum_{\bm{T}A} Q_{\bm{T}A,\bm{0}m}^{p-2} \left[\left(\Im P_{\bm{T}A,\bm{0}\bm{k}}\right)_{mn}\right]^2  \\
    \end{split}
    }
    \end{equation}

    \subsubsection{Connected asymmetric term}

    \begin{equation}
    \begin{split}
        D^{X}_{\bm{k}mn}
            &= -2p \sum_{\bm{T}A,i} Q_{\bm{T}A,\bm{0}i}^{p-1} \Re \left(
                P_{\bm{T}A,\bm{0}\bm{k}} \frac{\partial^2 U_{\bm{k}}}{\partial X_{\bm{k}mn}^2}
            \right)_{ii}    \\
            &= -2p \sum_{\bm{T}A,i} Q_{\bm{T}A,\bm{0}i}^{p-1} \Re \left[
                P_{\bm{T}A,\bm{0}\bm{k}} (E_{mn} - E_{nm})^2
            \right]_{ii}    \\
            &= 2p \sum_{\bm{T}A,i} Q_{\bm{T}A,\bm{0}i}^{p-1} \Re \left( P_{\bm{T}A,\bm{0}\bm{k}} E_{mm} \right)_{ii}
            + (m \leftrightarrow n)    \\
    \end{split}
    \end{equation}
    \begin{equation}
    \begin{split}
        D^{Y}_{\bm{k}mn}
            &= -2p \sum_{\bm{T}A,i} Q_{\bm{T}A,\bm{0}i}^{p-1} \Re \left(
                P_{\bm{T}A,\bm{0}\bm{k}} \frac{\partial^2 U_{\bm{k}}}{\partial Y_{\bm{k}mn}^2}
            \right)_{ii}    \\
            &= 2p \sum_{\bm{T}A,i} Q_{\bm{T}A,\bm{0}i}^{p-1} \Re \left[
                P_{\bm{T}A,\bm{0}\bm{k}} (E_{mn} + E_{nm})^2
            \right]_{ii}    \\
            &= 2p \sum_{\bm{T}A,i} Q_{\bm{T}A,\bm{0}i}^{p-1} \Re \left( P_{\bm{T}A,\bm{0}\bm{k}} E_{mm} \right)_{ii}
            + (m \leftrightarrow n)    \\
    \end{split}
    \end{equation}
    Therefore
    \begin{equation}
    \boxed{
        \tilde{D}^{X}_{\bm{k}mn}
            = \tilde{D}^{Y}_{\bm{k}mn}
            = 2p \sum_{\bm{T}A} Q_{\bm{T}A,\bm{0}m}^{p-1} \left(\Re P_{\bm{T}A,\bm{0}\bm{k}}\right)_{mm}
    }
    \end{equation}

    \subsubsection{Connected symmetric term}

    \begin{equation}
    \begin{split}
        D^{X}_{\bm{k}mn}
            &= -2p \sum_{\bm{T}A,i} Q_{\bm{T}A,\bm{0}i}^{p-1} \Re \left(
                \frac{\partial U_{\bm{k}}^{\dagger}}{\partial X_{\bm{k}mn}} P_{\bm{T}A,\bm{k}\bm{k}}
                \frac{\partial U_{\bm{k}}}{\partial X_{\bm{k}mn}}
            \right)_{ii}    \\
            &= 2p \sum_{\bm{T}A,i} Q_{\bm{T}A,\bm{0}i}^{p-1} \Re \left[
                (E_{mn} - E_{nm}) P_{\bm{T}A,\bm{k}\bm{k}} (E_{mn} - E_{nm})
            \right]_{ii}    \\
            &= -2p \sum_{\bm{T}A,i} Q_{\bm{T}A,\bm{0}i}^{p-1} \Re \left(
                E_{mn} P_{\bm{T}A,\bm{k}\bm{k}} E_{nm}
            \right)_{ii} + (m \leftrightarrow n)
    \end{split}
    \end{equation}
    \begin{equation}
    \begin{split}
        D^{Y}_{\bm{k}mn}
            &= -2p \sum_{\bm{T}A,i} Q_{\bm{T}A,\bm{0}i}^{p-1} \Re \left(
                \frac{\partial U_{\bm{k}}^{\dagger}}{\partial Y_{\bm{k}mn}} P_{\bm{T}A,\bm{k}\bm{k}}
                \frac{\partial U_{\bm{k}}}{\partial Y_{\bm{k}mn}}
            \right)_{ii}    \\
            &= -2p \sum_{\bm{T}A,i} Q_{\bm{T}A,\bm{0}i}^{p-1} \Re \left[
                (E_{mn} + E_{nm}) P_{\bm{T}A,\bm{k}\bm{k}} (E_{mn} + E_{nm})
            \right]_{ii}    \\
            &= -2p \sum_{\bm{T}A,i} Q_{\bm{T}A,\bm{0}i}^{p-1} \Re \left(
                E_{mn} P_{\bm{T}A,\bm{k}\bm{k}} E_{nm}
            \right)_{ii} + (m \leftrightarrow n)
    \end{split}
    \end{equation}
    Therefore
    \begin{equation}
    \boxed{
        \tilde{D}^{X}_{\bm{k}mn}
            = \tilde{D}^{Y}_{\bm{k}mn}
            = -2p \sum_{\bm{T}A} Q_{\bm{T}A,\bm{0}m}^{p-1} \left(
                \Re P_{\bm{T}A,\bm{k}\bm{k}}
            \right)_{nn}
    }
    \end{equation}

    \section{Molecular CIAH-based PM optimization}


    The CIAH-based PM optimization for molecules and periodic solids in the $\Gamma$-point supercell formulation, as previously implemented in PySCF~\cite{Sun16arXiv}, is computationally dominated by the Hessian--vector product,
    \begin{equation}
        (\mathbf{Hv})_{ij}
            = f_{ij} \hat{A}_{ij} \{
                \tilde{\sigma}^{\mathrm{d}}_{ij}
                + \tilde{\sigma}^{\textrm{c-symm}}_{ij}
                + \tilde{\sigma}^{\textrm{c-asymm}}_{ij}
            \},
    \end{equation}
    where the disconnected term is
    \begin{equation}
        \tilde{\sigma}^{\mathrm{d}}_{ij}
            = - 4 p(p-1) \sum_{A}^{N_{\mathrm{atom}}} Q_{A,j}^{p-2} \Re(v^{\dagger} P_A)_{jj} P_{A,ij},
    \end{equation}
    the connected symmetric term is
    \begin{equation}
        \tilde{\sigma}^{\textrm{c-symm}}_{ij}
            = -2p \sum_{A}^{N_{\mathrm{atom}}} Q_{A,j}^{p-1} \left(P_{A} v\right)_{ij},
    \end{equation}
    and the connected asymmetric term is
    \begin{equation}
        \tilde{\sigma}^{\textrm{c-asymm}}_{ij}
            = - p \sum_{m}^{N_{\mathrm{orb}}} \left(
                \sum_{A} Q_{A,m}^{p-1} P_{A,im} \right) v^*_{jm}
            - p \sum_{A}^{N_{\mathrm{atom}}} Q_{A,j}^{p-1} (v^{\dagger} P_{A})_{ij}.
    \end{equation}
    Here,
    \begin{equation}    \label{eq:mol_ciah_PAij}
        P_{A,ij}
            = \braket{w_{i} | \hat{P}_A | w_{j}}
            = \sum_{\mu \in A} O_{\mu i}^* O_{\mu j}
    \end{equation}
    is the projection matrix, and $Q_{A,i} = P_{A,ii}$ is the atomic population associated with localized orbital $i$.
    In \cref{eq:mol_ciah_PAij}, $O_{\mu i}$ denotes the overlap between the projectors and the localized orbitals, which can be evaluated from their AO coefficient matrices as
    \begin{equation}
        O_{\mu i}
            = \braket{ \chi_{\mu} | w_{i} }
            = \sum_{\rho \tau}^{N_{\mathrm{AO}}^2} D_{\rho \mu}^* S_{\rho\tau} C_{\tau i}.
    \end{equation}
    When the molecular CIAH algorithm is applied to a periodic supercell with $\Gamma$-point sampling, all indices are understood as supercell indices.
    Accordingly, $N_{\mathrm{atom}} = N_k n_{\mathrm{atom}}$, $N_{\mathrm{orb}} = N_{\mathrm{k}} n_{\mathrm{orb}}$, $N_{\mathrm{proj}} = N_{\mathrm{k}} n_{\mathrm{proj}}$, and $N_{\mathrm{AO}} = N_{\mathrm{k}} n_{\mathrm{AO}}$.

    As written, the CPU cost of the Hessian--vector product is dominated by forming the projection--vector product,
    \begin{equation}    \label{eq:Pv_quartic}
        (P_A v)_{ij}
            = \sum_{m}^{N_{\mathrm{orb}}} P_{A, im} v_{mj},
    \end{equation}
    which scales as $O(N_{\mathrm{atom}} N_{\mathrm{orb}}^3) = O(N_{k}^4 n_{\mathrm{atom}} n_{\mathrm{orb}}^3) \sim O(N_k^4 n^4)$.
    In this work, we employ an optimization analogous to that described in Section \fakeref{IIE} of the main text to improve the computational efficiency of molecular CIAH.
    In particular, the quartic scaling of \cref{eq:Pv_quartic} can be reduced to cubic by exploiting the factorized form of the projection matrix in \cref{eq:mol_ciah_PAij},
    \begin{equation}
        (P_A v)_{ij}
            = \sum_{\mu \in A} O_{\mu i}^* \left(
                \sum_{m}^{N_{\mathrm{orb}}} O_{\mu m} v_{mj}
            \right),
    \end{equation}
    which can be evaluated in two steps, each scaling as $O(N_{\mathrm{proj}} N_{\mathrm{orb}}^2) = O(N_k^3 n_{\mathrm{proj}} n_{\mathrm{orb}}^2) \sim O(N_k^3 n^3)$.

    The memory cost is dominated by storage of the projection matrix $P_{A,ij}$, whose size scales as $O(N_{\mathrm{atom}} N_{\mathrm{orb}}^2) \sim O(N_k^3 n^3)$.
    In principle, this cost could be reduced to $O(N_k^2 n^2)$ by using \cref{eq:mol_ciah_PAij} directly whenever $P_{A,ij}$ is required.
    In practice, however, we retain the explicit $P_{A,ij}$ tensor, which substantially simplifies the evaluation of the gradient and the remaining contributions to the Hessian--vector product.

    \section{Jacobi sweep}

    \subsection{General consideration}

    A general real-valued, translationally symmetric Jacobi rotation can be written as
    \begin{equation}
        \label{eq:Jacobi_real_space}
        \begin{bmatrix}
            \tilde{w}_{\bm{R}_0, i}  &
            \tilde{w}_{\bm{R}_0+\bm{R}, j}
        \end{bmatrix}
        =
        \begin{bmatrix}
            w_{\bm{R}_0, i}  &
            w_{\bm{R}_0+\bm{R}, j}
        \end{bmatrix}
        \begin{bmatrix}
            \cos\theta & \sin\theta \\
            -\sin\theta & \cos\theta
        \end{bmatrix},
    \end{equation}
    which mixes $w_{\bm{0},i}$ with $w_{\bm{R},j}$ and, by translational covariance, applies the same rotation to all their lattice translates.
    Fourier transforming \cref{eq:Jacobi_real_space} yields the corresponding $k$-space formulation,
    \begin{equation}    \label{eq:Jacobi_k_space}
        \begin{bmatrix}
            \tilde{\phi}_{\bm{k} i}  &
            \tilde{\phi}_{\bm{k} j}
        \end{bmatrix}
        =
        \begin{bmatrix}
            \phi_{\bm{k} i}  &
            \phi_{\bm{k} j}
        \end{bmatrix}
        \begin{bmatrix}
            \cos\theta & \me^{\mi\bm{k}\cdot\bm{R}}\sin\theta \\
            -\me^{-\mi\bm{k}\cdot\bm{R}}\sin\theta & \cos\theta
        \end{bmatrix},
    \end{equation}
    which corresponds to a \emph{$k$-dependent} $2\times2$ rotation between Bloch orbitals $i$ and $j$.

    \subsection{Working equations for $p = 2$}

    To determine the optimal rotation angle $\theta$, consider the PM objective for a pair of WFs $(\bm{0}i,\bm{R}j)$ after applying the Jacobi rotation in \cref{eq:Jacobi_real_space}:
    \begin{equation}
        \tilde{L}_{ij}
            = \sum_{\bm{T}A}
            \big[(\tilde{P}_{\bm{T}A,\bm{0}\bm{0}})_{ii}^2
            + (\tilde{P}_{\bm{T}A,\bm{0}\bm{0}})_{jj}^2\big].
    \end{equation}
    Using
    \begin{equation}
    \begin{split}
        \tilde{w}_{\bm{0}i}
            &= w_{\bm{0}i} \cos\theta - w_{\bm{R}j} \sin\theta,  \\
        \tilde{w}_{\bm{0}j}
            &= w_{-\bm{R}i} \sin\theta + w_{\bm{0}j} \cos\theta,
    \end{split}
    \end{equation}
    the diagonal projection matrix elements after rotation become
    \begin{equation}
    \begin{split}
        (\tilde{P}_{\bm{T}A,\bm{0}\bm{0}})_{ii}
            &= (P_{\bm{T}A,\bm{0}\bm{0}})_{ii} \cos^2\theta
            + (P_{\bm{T}A,\bm{R}\bm{R}})_{jj} \sin^2\theta
            - 2\,\Re (P_{\bm{T}A,\bm{0}\bm{R}})_{ij} \cos\theta \sin\theta,    \\
        (\tilde{P}_{\bm{T}A,\bm{0}\bm{0}})_{jj}
            &= (P_{\bm{T}A,(-\bm{R})(-\bm{R})})_{ii} \sin^2\theta
            + (P_{\bm{T}A,\bm{0}\bm{0}})_{jj} \cos^2\theta
            + 2\,\Re (P_{\bm{T}A,(-\bm{R})\bm{0}})_{ij} \cos\theta \sin\theta \\
            &= (P_{\bm{T}A,\bm{0}\bm{0}})_{ii} \sin^2\theta
            + (P_{\bm{T}A,\bm{R}\bm{R}})_{jj} \cos^2\theta
            + 2\,\Re (P_{\bm{T}A,\bm{0}\bm{R}})_{ij} \cos\theta \sin\theta.
    \end{split}
    \end{equation}
    In the second line for $(\tilde{P}_{\bm{T}A,\bm{0}\bm{0}})_{jj}$, we have used translational symmetry together with the fact that $\bm{T}$ is summed over in the final objective.
    The change in the PM objective due to the $(\bm{0}i,\bm{R}j)$ rotation is therefore
    \begin{equation}
        \label{eq:Jacobi_Delta_Lij}
        \Delta L_{ij}(\theta)
            = \tilde{L}_{ij}(\theta) - L_{ij}
            = -\frac{1}{2} \sin2\theta \left[
                (A_{\bm{0}\bm{R}})_{ij} \cos2\theta
                + (B_{\bm{0}\bm{R}})_{ij} \sin2\theta
            \right],
    \end{equation}
    where we have introduced the intermediates
    \begin{equation}
    \begin{split}
        (A_{\bm{0}\bm{R}})_{ij}
            &= 4 \sum_{\bm{T}A} \big[(P_{\bm{T}A,\bm{0}\bm{0}})_{ii} - (P_{\bm{T}A,\bm{R}\bm{R}})_{jj}\big] \Re (P_{\bm{T}A,\bm{0}\bm{R}})_{ij},   \\
        (B_{\bm{0}\bm{R}})_{ij}
            &= \sum_{\bm{T}A} \big[(P_{\bm{T}A,\bm{0}\bm{0}})_{ii} - (P_{\bm{T}A,\bm{R}\bm{R}})_{jj}\big]^2 - [2\,\Re (P_{\bm{T}A,\bm{0}\bm{R}})_{ij}]^2.
    \end{split}
    \end{equation}
    Setting $\Delta L_{ij}'(\theta)=0$ yields
    \begin{equation}
        \label{eq:Jacobi_theta_eqn}
        \tan 4\theta
            = -\frac{
                (A_{\bm{0}\bm{R}})_{ij}
            }{
                (B_{\bm{0}\bm{R}})_{ij}
            }.
    \end{equation}
    \Cref{eq:Jacobi_theta_eqn} can be interpreted as follows:
    \begin{enumerate}
        \item Since the input orbitals are already at a stationary point of the PM functional, $\theta=0$ is a trivial solution of \cref{eq:Jacobi_theta_eqn}, implying that the right-hand side vanishes. The non-trivial solutions therefore satisfy $\tan 4\theta = 0$, i.e., $4\theta = n\pi$ with $n\neq 0$.
        \item Furthermore, it is sufficient to restrict $\theta$ to the interval $[0,\pi)$, because a rotation by $\theta=\pi$ corresponds to a simultaneous sign change of the two orbitals, which leaves the PM objective invariant, and any $\theta>\pi$ can be mapped back into $[0,\pi)$.
    \end{enumerate}
    We therefore conclude that the only distinct non-trivial solutions are
    \begin{equation}
        \label{eq:Jacobi_theta_solution}
        \theta = \frac{\pi}{4},\; \frac{\pi}{2},\; \frac{3\pi}{4}.
    \end{equation}
    In practice, we can calculate $\Delta L_{ij}$ for all possible $\theta$ values in \cref{eq:Jacobi_theta_solution} to find one that maximizes the PM objective.

    \subsection{Extension for $p > 2$}

    The derivation above strictly assumes $p=2$.
    A similar derivation for $p = 3$ leads to the same condition \cref{eq:Jacobi_theta_eqn} for determining $\theta$ except that the two intermediates are defined as
    \begin{equation}
    \begin{split}
        (A_{\bm{0}\bm{R}})_{ij}
            &= 4\sum_{\bm{T}A} \left[
                (P_{\bm{T}A,\bm{0}\bm{R}})_{ii}^2 - (P_{\bm{T}A,\bm{0}\bm{R}})_{jj}^2
            \right] \Re (P_{\bm{T}A,\bm{0}\bm{R}})_{ij},  \\
        (B_{\bm{0}\bm{R}})_{ij}
            &= \sum_{\bm{T}A} \left[
                (P_{\bm{T}A,\bm{0}\bm{R}})_{ii} + (P_{\bm{T}A,\bm{0}\bm{R}})_{jj}
            \right] \left\{
                \left[
                    (P_{\bm{T}A,\bm{0}\bm{R}})_{ii} - (P_{\bm{T}A,\bm{0}\bm{R}})_{jj}
                \right]^2 - \left[ 2\Re (P_{\bm{T}A,\bm{0}\bm{R}})_{ij} \right]^2
            \right\},  \\
    \end{split}
    \end{equation}
    For general $p>3$, an analytic derivation becomes cumbersome. However, the second argument above (periodicity in $\theta$) always holds, and we have numerically validated that the first argument holds for $p \leq 6$ (which arguably already covers most use cases of PM localization).
    We therefore \emph{postulate} that the first argument remains valid for all $p \geq 2$.

    \subsection{Efficient implementation and computational cost}

    In practice, we evaluate $\Delta L_{ij}(\theta)$ numerically for the three candidate angles in \cref{eq:Jacobi_theta_solution} and check whether any of them increases the PM objective.
    The computational cost of this Jacobi stability check scales as
    \begin{equation}
        O(N_k^2\,n_{\textrm{proj}}\,n_{\textrm{orb}}^2)
    \end{equation}
    for building the $(P_{\bm{T}A,\bm{k}\bm{R}})_{ij}$ intermediate.
    This cost is modest and comparable to that of a single gradient evaluation.

    However, one can reasonably expect that pairwise instabilities primarily involve orbital pairs that are not too far apart in real space.
    We therefore can optionally restrict the real-space shifts in \cref{eq:Jacobi_k_space} to satisfy
    \begin{equation}
        \label{eq:Jacobi_R_leq_Rmax}
        |\bm{R}| < R_{\textrm{max}},
    \end{equation}
    for a chosen cutoff $R_{\textrm{max}}$, which provides a controllable reduction in computational cost.
    Let $N_R$ denote the number of lattice vectors $\bm{R}$ satisfying \cref{eq:Jacobi_R_leq_Rmax}.
    Under this restriction, the cost of the Jacobi stability check becomes
    \begin{equation}
        O(N_k\,N_{R}\,n_{\textrm{proj}}\,n_{\textrm{orb}}^2),
    \end{equation}
    which is linear in $N_k$ when $N_R$ is independent of $N_k$.
    We choose $R_{\textrm{max}} = 10$~Bohr in this work, which covers WF pairs that are separated by roughly $5$ chemical bonds.

    \bibliography{refs_si}